\documentclass[11pt,letterpaper]{article}
\pdfoutput=1

\usepackage{caption}
\usepackage{graphicx,array}
\usepackage[dvipsnames,table]{xcolor}
\usepackage{mathrsfs}

\definecolor{darkblue}{rgb}{0,0.1,0.5}
\definecolor{darkgreen}{rgb}{0,0.5,0.2}
\definecolor{darkred}{RGB}{153,26,0}
\definecolor{seablue}{rgb}{0,0.2,0.6}
\definecolor{light}{rgb}{0,0.2,0}
\definecolor{viola}{RGB}{134,41,198}
\definecolor{myG}{RGB}{20,140,0}

\usepackage{slashed}
\usepackage{latexsym}
\usepackage{amssymb, amsmath,tabularx}
\usepackage{braket}
\usepackage{bm}
\usepackage{pifont}
\newcommand{\cmark}{\ding{51}}%
\newcommand{\xmark}{\ding{55}}%

\usepackage[numbers,sort&compress]{natbib}

\usepackage{mathrsfs}
\usepackage{adjustbox}
\usepackage{placeins}
\usepackage{afterpage}
\usepackage{hyperref} 
\hypersetup{
    colorlinks=true,       
    linkcolor=darkblue,          
    citecolor=BrickRed,        
    filecolor=magenta,      
    urlcolor=MidnightBlue           
}
\usepackage[all]{hypcap} 
\usepackage{subcaption}
\usepackage[font=small,labelfont=bf,labelsep=period]{caption}
\usepackage{tikz}
\usepackage{tkz-euclide}
\usetikzlibrary{shapes.misc}
\usetikzlibrary{decorations.pathreplacing,calligraphy}
\usetikzlibrary{decorations.pathmorphing} 
\tikzset{
	v/.style={decorate, decoration={snake, segment length=3mm, amplitude=0.75mm}, draw},
	f/.style={draw=black, postaction={decorate},
		decoration={markings,mark=at position .6 with {\arrow[very thick]{latex}}}},
	fb/.style={draw=black, postaction={decorate},
		decoration={markings,mark=at position .4 with {\arrowreversed[very thick]{latex}}}},
	fnar/.style={draw=black},
	g/.style={decorate, draw=black,
		decoration={coil,amplitude=3pt, segment length=3.5pt}},
	s/.style={dashed,draw=black, postaction={decorate},
		decoration={markings,mark=at position .55 with {\arrow[very thick]{latex}}}},
	sb/.style={dashed,draw=black, postaction={decorate},
		decoration={markings,mark=at position .55 with {\arrowreversed[draw=black,very thick]{latex}}}},
	snar/.style={dashed,draw=black,line width =1.5pt},
	cross/.style={cross out, draw=black, minimum size=2*(#1-\pgflinewidth), inner sep=0pt, outer sep=0pt},
	cross/.default={3pt},
	none/.style={draw=white, postaction={decorate},
		decoration={markings,mark=at position .6 with {\arrow[very thick]{latex}}}}
		}

\newcommand{\Sss}{\mathcal{S}_{s's}}
\newcommand{\IV}{\mathcal{I}_V}

\newcommand{\SV}{\vec S_V}

\newcommand{\Lag}{\mathscr{L}}
\newcommand{\Op}{\mathcal{O}}
\newcommand{\wc}{\mathcal{C}}
\newcommand{\M}{\mathcal{M}}
\newcommand{\FS}{V_{\mu\nu}\bar V^{\mu\nu}}
\newcommand{\OFS}{\tilde V_{\mu\nu}\bar V^{\mu\nu}}

\newcommand{\vev}{\mathrm{v}}

\newcommand{\GeV}{\mathrm{GeV}}

\newcommand{\ie}{\textit{i.e. }}
\newcommand{\eg}{\textit{e.g. }}

\newcommand{\lsix}{\widehat{\mathscr{L}}_{\rm SM,6}}
\newcommand{\leff}{\widehat{\mathscr{L}}_{\rm eff}}

\newcommand{\UD}{U(1)_\textup{\tiny{D}}}
\newcommand{\gd}{g_\textup{\tiny{D}}}
\newcommand{\gdp}{g^\prime_\textup{\tiny{D}}}

\newcommand{\OS}{\mathcal{O}_S}

\newcommand{\be}{\begin{equation}}
\newcommand{\ee}{\end{equation}}

\begin{document}

\begin{flushright}

\end{flushright}
\vspace{.6cm}
\begin{center}
{\LARGE \bf 
Complex Dark Photon Dark Matter EFT
}\\
\bigskip\vspace{1cm}
{
\large Enrico Bertuzzo$^{a,b,c}$, Tommaso Sassi$^{d,e}$, Andrea Tesi$^f$}
\\[7mm]
 {\it \small
$^a$Dipartimento di Scienze Fisiche, Informatiche e Matematiche,\\
Universit\`a degli Studi di Modena e Reggio Emilia, Via Campi 213/A, I-41125 Modena, Italy\\
$^b$INFN sezione di Bologna, via
Irnerio 46, 40126 Bologna, Italy\\
$^c$Instituto de F\'isica, Universidade de S\~ao Paulo, C.P. 66.318, 05315-970 S\~ao Paulo, Brazil\\
$^d$Dipartimento di Fisica e Astronomia, Universit\`a degli Studi di Padova, Via Marzolo 8, 35131 Padova, Italy \\
$^e$INFN, Sezione di Padova, Via Marzolo 8, 35131 Padova, Italy\\
$^f$INFN Sezione di Firenze, Via G. Sansone 1, I-50019 Sesto Fiorentino, Italy}
\end{center}

\vspace{.2cm}

\begin{abstract}
We construct an effective field theory for complex Stueckelberg dark photon dark matter. Such an effective construction can be realized by writing down a complete set of operators up to dimension six built with the complex dark photon and Standard Model fields. Classifying the effective operators, we find that in order to properly take into account the non-renormalizable nature of an interacting massive vector, the size of the Wilson coefficients should be naturally smaller than naively expected. This can be consistently taken into account by a proper power counting, that we suggest. First we apply this to collider bounds on light dark matter, then to direct detection searches by extending the list of non-relativistic operators to include the case of complex vectors. In the former we correctly find scaling limits for small masses, while in the latter we mostly focus on electric dipole interactions, that are the signatures of this type of dark matter.  Simple UV completions that effectively realize the above scenarios are also outlined.
\end{abstract}

\vfill
\noindent\line(1,0){188}
{\scriptsize{ \\ E-mail:\texttt{  \href{mailto:enrico.bertuzzo@unimore.it}{enrico.bertuzzo@unimore.it}, \href{mailto:tommaso.sassi@phd.unipd.it}{tommaso.sassi@phd.unipd.it}, \href{andrea.tesi@fi.infn.it}{andrea.tesi@fi.infn.it}}}}

\newpage
\tableofcontents

\section{Introduction}
The existence of Dark Matter (DM) relies so far just on gravitational evidences, while little is known about its possibile non-gravitational interactions (see however \cite{Cirelli:2024ssz} for a review). From the particle physics point of view, apart from the obvious absence of strong interactions with us, one of the most remarkable inferred property of DM is its cosmological stability. For ordinary matter, stability arises as a consequence of accidental symmetries of the Standard Model (SM) after all the possible renormalizable interaction terms are written down. Such accidents depend upon the gauge interactions as well as the matter content of the theory. Barring massless particles, as the photons, the stability of electrons and protons are understood in terms of accidental symmetry. In the DM literature the stability is often taken for granted and ensured, for example for phenomenological studies, assuming DM to be (at least) charged under some $\mathbb{Z}_2$ symmetry, with the SM being a total singlet. Generally speaking, such dark symmetries like $\mathbb{Z}_2$ can act on DM, so that in these cases it is then possible to study in full generality the effective theory of DM interactions with the SM \cite{Criado:2021trs,deBlas:2017xtg}.

In this work, we would like to study effective field theories (EFTs) of DM that are consistent with the idea that DM stability arises accidentally. As such - in analogy to what happens for the SM - we have in mind dark sectors with gauge interactions, where the lightest states are accidentally stable against decay to the SM.

Here we consider DM as a massive vector, a scenario which is commonly known as dark photon, see \cite{Fabbrichesi:2020aa,Caputo:2021eaa} for reviews.
Its simplest realization is characterised by a kinetic mixing interaction with the SM hypercharge. Dark photon can be DM if a $\mathbb{Z}_2$ symmetry forbids such a kinetic mixing 
\cite{Lebedev:2011iq,Farzan:2012hh,Baek:2012se,Duch:2015jta,DiFranzo:2015nli,Duch:2017khv,Azevedo:2018oxv,YaserAyazi:2019caf,Glaus:2019itb,Arcadi:2020jqf,Ghorbani:2021yiw,Arcadi:2021mag,Amiri:2022cbv,Frandsen:2022klh}. More complicated models involving  extensions of the SM with non abelian gauge groups have also been considered, realising stable vector DM candidates by means of renormalisable SU(2) gauge interactions \cite{Hambye:2008bq,Baek:2013dwa,Chen:2015dea,Kim:2015hda,Chen:2015nea,Khoze:2016zfi,Barman:2017yzr,Barman:2018esi,Chaffey:2019fec,Nomura:2020zlm,Baouche:2021wwa, Hu:2021pln,Arcadi:2021mag,Hu:2021pln,Zhang:2022wek,Frandsen:2022klh,Belyaev:2022aa}
(see also \cite{Hambye:2009fg} for an example of confined SU(2)), 
or larger groups as in \cite{Karam:2016rsz,Ko:2016fcd,Arcadi:2016kmk,Arcadi:2021mag}, further generalized to SU($N$) in \cite{Gross:2015cwa,DiChiara:2015bua,Frigerio:2022kyu}. The case in which the dark sector is not a complete singlet under the SM gauge group has been investigated for instance in \cite{Saez:2018off,Abe:2020mph}, allowing for electroweak interactions between the two sectors. 


Here we focus on a scenario in which the complex dark photon stability arises thanks to an accidental $\UD$ global symmetry of the dark sector. In order to fix the ideas, it is convenient to think about the SM. If we consider a world where there are no fermions, the $W^\pm$ vector bosons will be accidentally stable thanks to a $U(1)$ symmetry (the global rephasing associated to electromagnetism), while the $Z$ boson (or the Higgs) have in principle no right to be stable. Indeed they would decay to $W^+W^-$ if $g$ were sufficiently small. From this very simple ideal example, we are led to consider dark sectors - endowed with dark gauge-Higgs interactions - that deliver dark photons in irreducible representation of $\UD$. Having that in mind, we will remain agnostic about the precise origin of the interactions in the dark sector, excluded for the presence of this $\UD$ accidental symmetry, and only make use of the assumption that the dark matter candidate is a massive complex dark photon that is a complete singlet with respect to SM interactions. We dub these as \textit{complex dark photon} scenarios.

The main purpose of this paper is to consider a bottom-up approach and construct the EFT for complex dark photons stabilized by the accidental $\UD$, for similar studies see \cite{Barducci:2021egn,Kribs:2022gri,Hisano:2010yh,Yu:2011by,Kumar:2015wya,Belyaev:2016pxe,Belyaev:2018pqr,Arcadi:2020jqf,Arcadi:2021mag,Aebischer:2022wnl}. We consider $M_\star$ as the effective mass scale of the new sector, while $M$ is the mass of DM, and explore contact interactions between operators constructed with dark photon fields and SM singlet operators. In this context, the interactions between the dark and SM sector are typically casted in terms of a \textit{portal} connecting them, often generated by contact terms of SM singlet fields or carried by some mediator of definite spin. The most studied scenario is that of a scalar portal \cite{Duch:2015jta,Lebedev:2011iq,Arcadi:2020jqf,Baek:2012se,Chen:2015dea,Amiri:2022cbv,Frandsen:2022klh,DiFranzo:2015nli,Choi:2017zww}, but also models involving a pseudo-scalar \cite{Kaneta:2017aa}, a fermion \cite{Fraser:2014yga,Belyaev:2022aa} or a vector \cite{Ko:2020qlt,Ko:2016fcd,Cai:2021wmu,Zhou:2022pom,Chiang:2013kqa,Davoudiasl:2013jma,Catena:2023use} have been considered.

At low energy, DM is described as a Stuckelberg massive vector. While this is a consistent theory when there are no interactions, the presence of interactions between DM and the SM 
challenges the construction of the EFT. In practice, as we will review below, depending on the strength of the DM-SM interactions, 
Stueckelberg massive vectors lose perturbativity around a scale $E \gg M$ that can be well below the natural EFT cutoff $M_\star$. 
To overcome this problem, in this work we suggest a modified version of the power counting for the size of the Wilson coefficients of our DM EFT that systematically include this effect, by associating to each non-derivative insertion of dark photon fields a suppression proportional to ratio of $M/M_\star$. This is instrumental to avoid non-physical effects that appear when computing observables with on-shell DM, such as collider bounds.

With the correct power counting, we then explore the direct detection and collider phenomenology (for other works along these lines with somewhat complementary studies, see \cite{Baek:2013dwa,Khoze:2014aa,Yu:2011by,Davoudiasl:2013jma,Choi:2017zww}). We highlight the importance of the Higgs portal even in the dark photon case and we emphasize that accidentally stable complex dark photon can have - in principle - dipole interactions that give rise to long-range effect in direct detection rates \cite{Chu:2023zbo, Hisano:2020qkq}. The latter effect is absent for real dark photons.

The paper is organized as follows. In section \ref{sec:CDP} we discuss the structure of the operators that can be written down for a complex dark photon, paying attention to redundancies in their definition and to the hidden cut-off inherent to the Stuckelberg nature of the DM. The setup is then applied to identify the leading terms in an EFT expansion in section \ref{sec:structure_EFT}. In section \ref{sec:matching} we move to lower energies and match the EFT obtained in the previous section into effective theories valid below the electroweak scale and at the nuclear scale. This matching is instrumental for correctly compute the limits on the new physics scale $M_\star$ coming from direct detection experiments. Section \ref{sec:pheno} constitutes the core of our phenomenological analysis, in which we show the limits on the parameter space coming from direct detection and colliders (focusing, in the later case, on Higgs and $Z$ boson decays), while in section \ref{sec:UV} we sketch two possible UV-complete theories 
that generate, at least in part, our low energy EFT. We draw our conclusions in section \ref{sec:conclusions} and relegate technical material to the appendices.

\section{Complex Dark Photon}\label{sec:CDP}
In this work we consider DM as a complex dark photon, that is a vector field $V_\mu$ with a non-trivial complex conjugate $\bar V_\mu$, being a total singlet of the SM. The Lagrangian of $V_\mu$ is described by the following terms
\begin{equation}\label{eq:basic}
\mathscr{L}_V = -\frac{1}{2}V_{\mu\nu}\bar V^{\mu\nu} + M^2 V_\mu \bar{V}^\mu + \mathscr{L}_{\rm interactions}(V_\mu;\mathrm{SM})\,\,.
\end{equation}
The free part of the above lagrangian displays a global symmetry $\UD$ under which $V_\mu \to e^{i \alpha} V_\mu$. Such a symmetry, when extended to the full lagrangian, including interactions, acts as a stabilising symmetry for DM, forbidding interactions between one single $V$ and SM fields, all taken to be neutral under $\UD$. Effectively, one can think of this model as if the dark photon $V_\mu$ were charged with some dark abelian charge. 

At this level, $V_\mu$ is a massive Proca field (gauge fixed Stueckelberg vector), described by the generalized Lorenz condition $\partial_\mu V^\mu=0$, such that it is the solution of a Klein-Gordon equation $(\Box +M^2)V_\mu=0$ with just three physical degrees of freedom. The Fourier modes in momentum$-p$ space are described by three polarization states $\varepsilon_\mu^\lambda(p)$ such that
$\sum_{\lambda} \varepsilon_\mu^\lambda(p) \varepsilon_\nu^{\lambda*}(p)= - g_{\mu\nu} +p_\mu p_\nu/M^2$.

We assume that, at low energies, $\UD$ is a good symmetry of the DM lagrangian. This greatly restricts the interactions between $V_\mu$ and the SM fields, allowing us to classify the possible structures arising at low energy in an EFT description.
The idea is that at some scale $M_\star$, new states coupled to $V_\mu$ are integrated out. In general, we expect the following structure
\be\label{eq:model}
\mathscr{L}=\mathscr{L}_{\rm SM} + \frac{1}{M_\star^2}\leff(\mathrm{SM}; V_\mu)-\frac{1}{2}V_{\mu\nu}\bar V^{\mu\nu} + M^2 V_\mu \bar{V}^\mu\,.
\ee
At this level, $\leff(\mathrm{SM}; V_\mu)$ contains all the interactions between DM and the SM, and we expect it to be a polynomial series in operators of higher dimension. The normalization is such that  $\leff(\mathrm{SM}; V_\mu)$ has dimension six, but this does not forbid that it can contain operators of any dimensions with the appropriate dimensionful Wilson coefficient. The above structure is meant to render explicit the decoupling limit when the mass scale $M_\star$ is taken arbitrarily large, a limit in which the DM becomes extremely weakly coupled to the SM. Clearly, our EFT is valid up to energies $E\lesssim M_\star$ and can only be applied in this range.

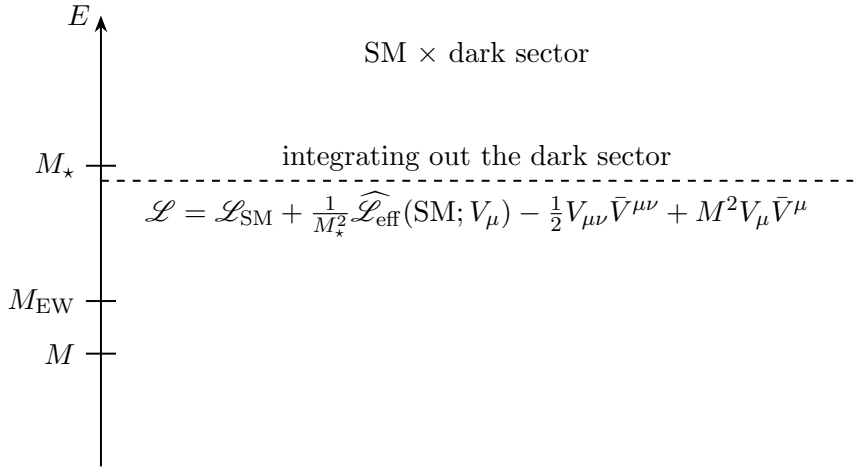
\begin{figure}[t]
\centering
\begin{tikzpicture}[line width=0.75]
\draw[-Stealth] (0,0) -- (0,6)node[left]{$E$};
\draw (-0.2,4)node[left]{$M_\star$} -- (0.2,4);
\draw[dashed] (0,3.8) -- (10,3.8)node[midway,above]{integrating out the dark sector}node[midway,below]{$\mathscr{L}=\mathscr{L}_{\rm SM} + \frac{1}{M_\star^2}\leff(\mathrm{SM}; V_\mu)-\frac{1}{2}V_{\mu\nu}\bar V^{\mu\nu} + M^2 V_\mu \bar{V}^\mu$};
\draw [decorate,
    decoration = {calligraphic brace}] (0.3,3.1) -- (0.3,1.3) node[midway, right]{$M$};
\draw[fill=gray] (-0.1, 1.3) rectangle ++(0.2, 1.8);
\draw (-0.2,2.2)node[left]{$M_\text{EW}$} -- (0.2,2.2);
\node[] at (5,5.5) {SM $\times$ dark sector};
\end{tikzpicture}
\caption{\label{fig:cartoon} Generation of the DM effective field theory under consideration. We put in evidence the hierarchy of scales needed for the effective description to be valid.}
\end{figure}

To be included in the EFT, the DM mass must satisfy  $M\ll M_\star$, so that all the process important for DM phenomenology can be described within the effective theory. The fundamental theory of which eq. \eqref{eq:basic} is a low-energy description is depicted in a cartoon in fig. \ref{fig:cartoon}. We take the heavy fields that are integrated out to have masses $M_\star\gg M_\textup{\tiny{EW}}$, with $M_\textup{\tiny{EW}}$ standing for the common mass scale of the heaviest SM states $(h,W,t)$. The matching is then done at a scale $M_\star$ where the SM electroweak symmetry is still unbroken and operators are classified according to SM gauge symmetries.

The systematic construction of the effective field theory coupling the SM and DM, described by $\leff(\mathrm{SM}; V_\mu)$, depends on two important assumptions that we now discuss. The first relies on the $\UD$ symmetry to be elevated to full symmetry of the lagrangians in eqs.\,\eqref{eq:basic} and\,\eqref{eq:model}. The second, instead, relies on the effective description of $V_\mu$ as a massive vector field \`a la Stueckelberg.
This implies that eq. \eqref{eq:basic} is itself an approximate description valid at low energies up to some scale $\Lambda_V$, that may be lower than $M_\star$. This last point is extremely important, as we can only use the EFT in a correct energy range where $\Lambda_V\gg M_\star$.

\subsection{Symmetry constraints on the effective interactions}\label{sec:operator_structures}
\begin{table}[tb]
    \centering
    \begin{tabular}{|>{\centering\arraybackslash}m{3cm}|>{\centering\arraybackslash}m{3cm}|>{\centering\arraybackslash}m{3cm}|}
        \hline
        Type & Name & Expression \\
        	\hline
	\hline
 	 \vspace{1mm}scalar & \vspace{1mm}$\Op_S$ & \vspace{1mm}$V_\mu \bar{V}^\mu$ \\
         scalar & $\Op_{F^2}$ & $V_{\mu\nu} \bar{V}^{\mu\nu}$ \\
       pseudoscalar \vspace{1mm}& $\Op_{P}$ \vspace{1mm}& $\varepsilon_{\mu\nu\rho\sigma}V^{\mu\nu}\bar V^{\rho\sigma}\vspace{1mm}$\\
       \hline
 	 vector & $J^\mu_V$ & \vspace{1mm}$i \bar{V}_\nu \overleftrightarrow{\partial_\mu} V^\nu$ \\
        pseudo-vector \vspace{1mm}& $J^\mu_P\vspace{1mm}$ & $ i\varepsilon^{\mu\nu\rho\sigma}\bar{V}_\nu \overleftrightarrow{\partial_\rho} V_\sigma\vspace{1mm}$ 		\\
	\hline	 
	symm. tensor &  \vspace{1mm}$\mathcal{O}^S_{\mu\nu}$ & \vspace{1mm}$V_{(\mu} \bar{V}_{\nu)}$ \\
        \vspace{1mm} antisymm. tensor & \vspace{1mm}$\mathcal{O}^A_{\mu\nu}$ & \vspace{1mm}$iV_{[\mu} \bar{V}_{\nu]} $ \\
         \vspace{1mm} 
	traceless tensor & $\mathcal{O}^T_{\mu\nu}$ & $iV^{}_{[\mu\rho} \bar{V}^\rho_{\,\,\,\nu]}$ \\
	\hline
        \end{tabular}
    \caption{The eight operators constructed with two complex dark photon dark matter fields considered in this work, including up to two derivatives.}
    \label{table:structures}
\end{table}

Since the full theory enjoys $\UD$ but the SM is neutral under such a symmetry, the effective interactions $\leff(\mathrm{SM}; V_\mu)$ are constructed with singlets of $\UD$, that necessarily involve at least two powers of $V_\mu$. This forbids for example a kinetic mixing between $V_{\mu\nu}$ and the hyper-charge field strength, which is usually the most dominant effect for the phenomenology of dark photons \cite{Fabbrichesi:2020aa,Hambye:2008bq}.

We classify the $\UD$ singlet operators in terms of their Lorentz structure. Since we are dealing with a model for DM we only consider operators involving up to two fields $V_\mu$, and up to dimension four. We now discuss in detail the different structures that may appear in the EFT operators, also shown in table \ref{table:structures}.

\paragraph{Scalar singlet operators}~\\
We first discuss Lorentz scalar operators constructed with $V_\mu$. There are both CP-even and CP-odd singlet operators:
\be
\mathrm{scalar:}\,\quad \OS= V_\mu \bar{V}^\mu\,, \quad \mathcal{O}_{F}= V_{\mu\nu}\bar V^{\mu\nu}\,\,; \quad\quad \mathrm{pseudoscalar:}\,\quad \mathcal{O}_{P}=\varepsilon_{\mu\nu\rho\sigma}V^{\mu\nu}\bar V^{\rho\sigma}\,.
\ee
Out of these three operators, 
$\OS=V_\mu \bar{V}^\mu$ 
is expected to give the larger effects for phenomenology, since it has the lowest dimensions. Notice also that $O_P$ could have been added to eq.\,\eqref{eq:basic}, but being a total derivative does not lead to effects at the perturbative level. 
As for $V_{\mu\nu}\bar V^{\mu\nu}$, it can appear in dimension six operators involving $|H|^2$ affecting, for example, Higgs decay. These operators can then be contracted with the following SM scalar singlet operators 
\be\label{eq:SM_scalar}
\mathcal{O}_{\rm SM} = \{\quad |H|^2\,,\quad |H|^4, \quad F^2, \quad |D_\mu H|^2\,, \quad \bar\psi i \slashed{D}\psi,  \quad Q H U,  \quad F \tilde F,\quad \cdots \quad \}\,.
\ee
Here $F=G,W,B$ refer to the SM gauge bosons in the unbroken electroweak phase, in agreement with the assumption $M_\star\gg M_\textup{\tiny{EW}}$.
In this paper we follow the convention by which $\langle H\rangle = \vev/\sqrt{2}(0,1)^T$ and $\vev\simeq 246\, \GeV$.

We notice that other Lorentz and $\UD$ invariant scalars can be constructed with $V_\mu$ and derivatives, such as $\partial_\mu V_\rho \partial^\mu \bar V^\rho$ or $|\partial_\rho V^\rho|^2$. However they are redundant with $V\bar V$ (and current operator, see next) once equations of motion and gauge conditions are imposed, for example $\partial_\mu V_\rho \partial^\mu \bar V^\rho=V_\rho \Box \bar V^\rho + \mathrm{total\ derivative}$. In addition, there is also the dimension-4 structure $\partial_\mu V^\nu \partial_\nu \bar{V}^\mu$, which after double integration by part can be written as a (symmetric) tensor structure, also to be discussed next.

\paragraph{Vector singlet operators}~\\
Lorentz vector operators are again of two types, depending on the CP quantum numbers. In particular we can define a current real CP-even operator and a CP-odd vector exploiting the Levi-Civita tensor
\be
\mathrm{vector:}\,\quad J_\mu^V=i \bar{V}_\nu \overleftrightarrow{\partial_\mu} V^\nu\,\,; \quad\quad \mathrm{pseudo-vector:}\,\quad J_\mu^P= i\varepsilon_{\mu\nu\rho\sigma}\bar{V}^\nu \overleftrightarrow{\partial^\rho} V^\sigma.
\ee

Notice that other CP-even structures such as $\bar V^\rho V_{\rho\mu}$  are equivalent to $J_\mu^V$ when integration by part and the condition $\partial_\mu V^\mu=0$ are taken into account.
These operators have energy dimension three and can be contracted with vector singlet operators of the SM, such as the hyper-charge current or generic vector structures bilinear in the SM fermions, Higgs and gauge fields. These operators can then be paired with the following vector structures of SM fields:
\be\label{eq:SM_vector}
\mathcal{O}_{\rm SM}^{\mu} = \{\quad J_Y^\mu \equiv \frac{\partial^\rho B^\mu_{\phantom{\mu}\rho}}{g'}\,,\quad J_H^\mu \equiv i H^\dag \overleftrightarrow{D^\mu} H, \quad \bar\psi_i \gamma^\mu \psi_i\,,\quad \cdots\quad \}\,,
\ee
where $\psi_i$ denotes a generic SM fermion with flavor index $i$. 
For later convenience we have introduced the hyper-charge current $J_Y^\mu$ and the Higgs singlet current $J_H^\mu$. We take the complex dark photon to have diagonal couplings to SM fermions, avoiding flavor-violation effects, from which follow equal flavor indices in the fermionic current.

\paragraph{Tensor singlet operators}~\\
In principle we can also classify tensor structures and distinguish them upon symmetry properties. In particular we can have both antisymmetric and symmetric combinations:
\be
\mathrm{antisymm.:}\,\quad \mathcal{O}^A_{\mu\nu}=iV_{[\mu} \bar{V}_{\nu]}\,,\quad \Op_{\mu\nu}^T= i V^{}_{[\mu\rho} \bar V^{\rho}_{\,\,\,\nu]}\,;\quad\quad \mathrm{symm.:}\,\quad \mathcal{O}^S_{\mu\nu}=V_{(\mu} \bar{V}_{\nu)} .
\ee
All these dark operators can be contracted with SM Lorentz tensor structures (also respecting gauge invariance), but notice that if contracted with terms proportional to $g_{\mu\nu}$ the resulting interactions are either redundant with the ones associated to scalar-scalar contact operators or vanish.
The $\Op^A_{\mu\nu}$ and $\Op^S_{\mu\nu}$ terms are dimension 2, but only the antisymmetric combination allows for the creation of a dimension-4 renormalizable interaction when coupled to $B_{\mu\nu}$, therefore resulting of particular interest since it provides leading phenomenological contributions. Meanwhile $\Op_{\mu\nu}^T$, and $\Op^S_{\mu\nu}$ when contracted with tensor structure with SM fields gives rise to higher dimensional effective operators. We remind that we do not consider operators in $\leff(\mathrm{SM}; V_\mu)$ with dimension higher than 6.

Having this in mind, dark tensor singlets can be coupled to the following SM tensor and gauge singlets:
\be\label{eq:SM_tensor}
\mathcal{O}_{\rm SM}^{\mu\nu} = \{\, B^{\mu\nu}\,,\,\,\, |H|^2B^{\mu\nu}\,,\,\,\,\tilde B^{\mu\nu}\,,\,\,\, |H|^2\tilde B^{\mu\nu}\,,\,\,\, \partial^{(\mu}\partial^{\nu)}|H|^2\,,\,\,\, T^{\mu\nu}_\psi\,,\,\,\, T_F^{\mu\nu}\,\}\,.
\ee
We here accounted for the possibility of giving rise to both CP-even and CP-odd interactions, where the couplings to the hodge dual $\tilde B^{\mu\nu}$ is the only extra source of CP-violation coming with SM structures. The tensor structure $T^{\mu\nu}_{\psi\,(F)}$ is nothing but the energy-momentum tensor of SM fermions (gauge bosons).

We summarize all the possible UV operator structures respecting symmetry constraints with two complex dark photons in table \ref{table:structures}.

\subsection{Stueckelberg effective field theory}\label{sec:stueckelberg}
An important consideration on the validity of the EFT is somehow already hidden in eq.\,\eqref{eq:basic} when a bare mass term for the vector has been introduced, with the same structure of the operator $\OS$. 

The free part of eq.\,\eqref{eq:basic} enjoys the typical structure of Stueckelberg massive vector, which in isolation is a renormalizable theory. However, interactions can easily spoil this behaviour, as we now show. This is crucially related to the operator $\OS$ which is badly behaved at high-energy when tested in connection with other interactions and it is of great importance for our effective description. 
Similar discussions along this line can be found in \cite{Kribs:2022gri} for a real massive Stuckelberg vector, and in \cite{Chu:2023zbo} for the case of complex vector only coupled to operators made with electromagnetic field strength. We generalize the discussion to an EFT approach. 

In order to illustrate the problem, let us consider as an example the operator $\kappa |H|^2 \OS$, which has exactly the same structure of the mass term, $M^2\OS$, of eq.\,\eqref{eq:basic}. Having added interactions, despite them being dimension-4 at this level, and hence \textit{naively} renormalizable, they completely spoil the argument that the Stueckelberg mass can be extrapolated up to extremely high energy. This follows from the fact that the amplitude for the scattering $hh\to V \bar V$ grows with energy for $E\gg M/\sqrt{\kappa}$, introducing a physical cut-off at a scale $\Lambda_V\approx M/\sqrt{\kappa}$ that, depending on the value of $\kappa$, can be parametrically smaller than $M_\star$.

The apparent renormalizability of $\kappa |H|^2 V_\mu \bar V^\mu$ is manifestly lost when, applying the `Stueckelberg trick,' we restore the gauge invariance of the free Proca Lagrangian. Introducing the Goldstone $\pi$ and by sending $V_\mu \to V_\mu + \partial_\mu \pi/M$, we restore the full gauge symmetry under which $\pi$ shifts. In particular the gauge invariance is realised as
\be\label{eq:gauge}
\delta V_\mu(x) =\partial_\mu \lambda(x)\,,\quad \quad \delta\pi(x)=-M\lambda(x)\,,
\ee
and complex conjugate transformations for $\bar{V}$ and $\bar{\pi}$.

This is instrumental to study the high energy behaviour of the model: when discussing the $E\gg M$ limit it is convenient to restore the gauge symmetry. This leads to the following Lagrangian from the free part of \eqref{eq:basic}
\be
\mathscr{L}_{V}^{\rm Stueckelberg}= -\frac12 V_{\mu\nu}\bar{V}^{\mu\nu} + M^2\left(V_\mu + \frac{\partial_\mu \pi}{M}\right)\left(\bar V^\mu + \frac{\partial^\mu \bar\pi}{M}\right) -\frac{1}{\xi}(\partial_\mu \bar V^\mu +\xi M \bar \pi)(\partial_\mu V^\mu + \xi M \pi)\,,
\ee
where we have added a $R_\xi$ gauge fixing to remove the $V-\pi$ mixing. In Landau gauge with $\xi=0$, the amplitudes with an external longitudinal polarization at high momentum $p\gg M$ are equivalent to $\mathcal{A}(V_L(p);...)=\varepsilon_L^\mu(p) \mathcal{A}_\mu=\mathcal{A}(\pi;...)$ as dictated by the `equivalence theorem'.

This can then be used to study the high energy behaviour of operators constructed with $V_\mu$. 
This feature does not appear in operators involving $V_{\mu\nu}$, 
since it is invariant under \eqref{eq:gauge} 
and the Goldstone boson $\pi$ does not appear. Each time an insertion of $V_\mu$ is present, the high energy behavior can be studied focusing on the longitudinal polarization of $V$ (in Landau gauge for simplicity). This has remarkable consequences for our DM EFT, for all kind of dark operator structures.

This can then be applied to study the cut-off associated to processes involving non gauge-invariant operators under the sole $\delta V_\mu =\partial_\mu \lambda$. Only operators constructed with $V_{\mu\nu}$ are gauge invariant, while $\OS$, $J_\mu^V,\,J_\mu^P$, $\Op_{\mu\nu}^A$ and $\Op_{\mu\nu}^S$ are not.
This implies that operators involving such terms are originating from integrating out (at tree or loop level) dark sector states that are involved with the generation of the mass scale $M$. It also makes evident that only in the singular limit where the Wilson coefficients of these operators are exactly zero the gauge invariance is recovered in the effective operators. In such a limit a local $U(1)^2\times \mathbb{Z}_2$ is recovered.

For all processes at energies much above $M$, such as the case of $H$ and $Z$ decays into dark matter, the behavior of these operator is respectively
\be
\OS = \frac{1}{M^2} |\partial_\mu \pi|^2+\cdots\,, \quad J_\mu^V = \frac{i}{M^2}(\partial_\rho \bar{\pi} \overleftrightarrow{\partial_\mu}\partial^\rho \pi)+\cdots\,,\quad \mathcal{O}_{\mu\nu}=\frac{\partial_\mu \pi \partial_\nu \bar{\pi}}{M^2}+\cdots
\ee
This signals the appearance of a physical cut-off that can be parametrically smaller than $M_\star$ and it is related to the mass $M$ rather than to some UV parameter.

For the case of interest $|H|^2\OS$, which will be relevant for our phenomenological discussion, we then have
\be
\kappa |H|^2 \OS = \frac{\kappa}{M^2} |H|^2 (\partial_\mu \pi + \cdots)(\partial^\mu \bar \pi +\cdots)\,.
\ee
The operator has a hidden cut-off which is much smaller than naively expected and, as mentioned above, is of order $M/\sqrt{\kappa}$. At this point, our normalization of the $\leff$ term in \eqref{eq:model} is justified even for the operator $|H|^2\OS$ and it suggests that the correct power counting for $\kappa$ is of the form $\kappa\propto \lambda \times (M/M_\star)^2$. Here $\lambda$ is a (quartic) coupling and $M_\star$ the scale of the new-physics generating the effective interaction. The physical cut-off of the effective operator is now $M_\star/\sqrt{\lambda}$, which is correctly the scale of new-physics $M_\star$ times the effect of the coupling between the SM and DM sectors.

\section{Structure of the EFT}\label{sec:structure_EFT}
Having discussed the possible operators constructed with $V_\mu$ invariant under $\UD$ and the constraints on the size of the coupling due to the parametrization as a Stueckelberg massive vector, we are now in the position to write down the effective lagrangian of SM plus DM. As already mentioned, this lagrangian has to be invariant under SM$\times\UD$. At energies below $M_\star$ the SM is deformed by the following interactions
\be\label{eq:lagrangian}
\mathscr{L}=\mathscr{L}_{\rm SM} -\frac{1}{2}V_{\mu\nu}\bar V^{\mu\nu} + M^2 V_\mu \bar{V}^\mu  + \leff(\mathrm{SM};V_\mu) + \frac{1}{M_\star^2}\lsix\,.
\ee
We have included also the SM EFT at dimension-6, $\lsix$, because in concrete realizations we expect that deformations with only SM field  be generated by integrating out the dark sector at the scale $M_\star$. For the moment, we only consider $\leff$. 
The construction of the EFT can proceed simply by listing all the operators of a given dimension built from SM singlet operators and DM singlet operators, in order to have an EFT invariant under $\UD$. 
Therefore we have
\be\label{eq:EFT}
 \leff(\mathrm{SM};V_\mu)= \leff^{\, (4)}+ \frac{1}{M_\star^2} \leff^{\, (6)} +\cdots
\ee
The structure of these two terms is provided below and it is constructed using the operators in table \ref{table:structures} together with the SM structures listed in eqs.\,\eqref{eq:SM_scalar},\,\eqref{eq:SM_vector} and\,\eqref{eq:SM_tensor}.
 
\paragraph{Dimension-4 terms}~\\
There are only three operators at dimension-4, given by 
\be\label{eq:dim4}
\leff^{\, (4)} = \lambda_H \Op_S |H|^2 + \lambda_B \Op^A_{\mu\nu} B^{\mu\nu}+ \lambda_B^\prime \Op^A_{\mu\nu} \tilde B^{\mu\nu}\,.
\ee
While the first term is common to any theory of dark photon, and it is essentially an Higgs-portal interaction, the second and third contributions only arise for complex dark photon scenarios, since the anti-symmetric two index tensor vanishes for real vectors. In particular, we are interested in the phenomenology associated to the $\lambda_B$, $\lambda_B^\prime$ coefficients, which induce a low-energy interaction between DM and the photon -- despite the complex dark photon being electrically neutral. 
These interactions can give enhanced features at low momentum transferred in direct detection, equivalent to magnetic and electric dipole moment for spin-1 dark matter respectively \cite{Hisano:2010yh,Hisano:2020qkq,Chu:2023zbo}. Being dimension four, the operators in eq.\,\eqref{eq:dim4} are expected to be the leading effect, which should be even more important for strongly coupled DM as in \cite{Antipin:2015xia}. However, we will see that the appropriate power counting connected to the discussion in sec.\,\ref{sec:stueckelberg} will make the size of the Wilson coefficients effectively smaller.
\paragraph{Dimension-6 terms}~\\
There are clearly many more operators at dimension 6, given by the following Lagrangian: 
\be
\label{eq:dim6}
\begin{split}
\leff^{\, (6)}&= \Op_S\bigg[\wc_{H}|H|^4 +\wc_{DH}|D_\mu H|^2+\sum_\psi\wc_{H\psi}\, y_\psi\, \bar\psi H\psi+\sum_F\left(\wc_F F_{\mu\nu}F^{\mu\nu}+\wc_{\tilde F} F_{\mu\nu}\tilde F^{\mu\nu}\right)\bigg]\\ 
& \quad+\wc_{H,2}\Op_{F^2}|H|^2 +\wc_{H,3}\Op_{P}|H|^2\\
& \quad+ J_\mu^V \bigg[{\displaystyle{\sum_\psi}}\wc_\psi \bar\psi\gamma^\mu\psi+\wc_Y J_Y^\mu+\wc_{Hc}J_H^\mu\bigg]+J_\mu^P \bigg[{\displaystyle{\sum_\psi}}\wc_{\tilde \psi}\bar\psi\gamma^\mu\psi+\wc_{\tilde Y}J_Y^\mu+\wc_{\tilde H c}J_H^\mu\bigg]\\
& \quad+\Op^S_{\mu\nu} \bigg[\wc_{D2H}\partial^{(\mu}\partial^{\nu)}|H|^2+\sum_\psi\wc_{T_\psi}T_\psi^{\mu\nu}+\sum_q\wc_{T_F}T^{\mu\nu}_F\bigg] \,\\
& \quad+ \Op^A_{\mu\nu}\bigg[\wc_{HB} |H|^2B^{\mu\nu}+\sum_\psi\wc_{H\psi,2}\,y_\psi\,\bar\psi H\sigma^{\mu\nu}\psi\bigg]+\wc_{B}\Op_{\mu\nu}^TB^{\mu\nu}\\
& \quad+ \Op^A_{\mu\nu}\wc_{HB}^\prime |H|^2\tilde B^{\mu\nu}+\varepsilon^{\mu\nu\rho\sigma}\Op^A_{\mu\nu}\sum_\psi\wc_{H\psi,3}\,y_\psi\,\bar\psi H\sigma_{\rho\sigma}\psi+\wc_{B}^\prime\Op_{\mu\nu}^T\tilde B^{\mu\nu} \,\,.
\end{split}
\ee
We have grouped the various contact interaction accordingly to their Lorentz structure. We explicitly multiply every interaction involving one left handed and one right handed SM fermion by the corresponding Yukawa coupling, in such a way that Minimal Flavor Violation is respected\,\cite{DAmbrosio:2002vsn}. 

\subsection{Power counting possibilities}\label{sec:power_counting}
\begin{table}[tb]
    \centering
    {\def\arraystretch{1.25}
\begin{tabular}{|>{\centering\arraybackslash}m{2cm}|>{\centering\arraybackslash}m{3cm}|>{\centering\arraybackslash}m{3cm}|>{\centering\arraybackslash}m{3cm}|>{\centering\arraybackslash}m{3cm}|}
        \hline
         Operator & Expression & Gauge invariance & Naive power-counting & Improved power-counting \\
        	\hline\hline
 	 $\Op_S$ & $V_\mu \bar{V}^\mu$ & \xmark & 1 & $\times (M/M_\star)^2$\\
	 $\Op_{F^2}$ & $V_{\mu\nu} \bar{V}^{\mu\nu}$ &\cmark & 1 & $1$\\
	 $\Op_P$ & $\varepsilon_{\mu\nu\rho\sigma}V^{\mu\nu}\bar V^{\rho\sigma}$ &\cmark & 1 & $1$\\
	 $J_\mu^V$ & $i \bar{V}_\nu \overleftrightarrow{\partial_\mu} V^\nu$ &\xmark & 1 & $\times (M/M_\star)^2$\\
	 $J_\mu^P$ & $i\varepsilon_{\mu\nu\rho\sigma}\bar{V}^\nu \overleftrightarrow{\partial^\rho} V^\sigma$ &\xmark & 1 & $\times (M/M_\star)^2$\\
	 $ \Op^S_{\mu\nu}$ & $V_{(\mu} \bar{V}_{\nu)}$ & \xmark & 1 & $\times (M/M_\star)^2$\\
	 $ \Op^A_{\mu\nu}$ & $iV_{[\mu} \bar{V}_{\nu]}$ & \xmark & 1 & $\times (M/M_\star)^2$\\
	 $ \Op^T_{\mu\nu}$ & $iV^{}_{[\mu\rho} \bar{V}^\rho_{\,\,\,\nu]}$ & \cmark & 1 & $1$\\
	 \hline
       \end{tabular}
       }
    \caption{We summarize the effect of the application of the \textit{improved power counting} of eq.\,\eqref{eq:improved_PC} to all the possible DM structures listed in tab.\,\ref{table:structures}, highlighting when the extra $(M^2/M_\star^2)$ factor modifies the usual \textit{naive power counting} to restore the gauge invariance of the low energy theory for a massive Stueckelberg vector.}
    \label{table:power-counting}
\end{table}

We would now like to assign to the various coefficients a natural size that respects a good high-energy behavior of the theory, consistent with having massive gauge fields in the EFT. Exploiting the discussion of section \ref{sec:stueckelberg}, we can argue in favour of the following power counting: every vector field $V_\mu$ that appears without the full field strength is assigned a weight
\be\label{eq:improved_PC}
\begin{split}
V_\mu &\to \frac{M}{M_\star} V_\mu\,,
\end{split}
\quad\quad \text{improved power-counting}\,.
\ee
while operators built out of the DM field strength are left invariant.
In principle this rescaling has an overall $O(1)$ dimensionless coefficient that we fix to unity in the rest of the paper (see however section\,\ref{sec:UV}).
 
The rescaling of eq.\,\eqref{eq:improved_PC} has several effects. First of all, while it does not change the dimensionality of the operators, it makes their Wilson coefficient consistent with a high-energy limit of the massive gauge field, as discussed in section\,\ref{sec:stueckelberg}. Second, it will guarantee a correct high-energy behaviour, namely when the energy is in the range $M\ll E\ll M_\star$,  we will not see spurious non-decoupling effects, which is instrumental for a correct interpretations of the collider bounds. This has also an impact for the low energy physics related to DM scatterings (and, in general, also annihilations), since it naturally suppresses the size of the effects by the DM mass. Therefore, although less spectacular than the effects on high-energy observables $M\ll E$, there are important consequences also at low energies.

In terms of operators, this implies that the Wilson coefficients should be rescaled according to what shown in table \ref{table:power-counting}. In the table, we contrast the improved power-counting introduced in eq.\,\eqref{eq:improved_PC} with the naive power-counting in which the $V_\mu$ rescaling is not applied.

Moreover, in order to make a stronger connection with possible UV completion, we define the size of the Wilson coefficient $\wc_i$ normalized (in $\hbar$ counting) to a coupling $g_\star$. Every coefficient $\wc_i$ in eq.~\eqref{eq:EFT} therefore has its explicit expression in terms of $g_\star$. The dictionary is the following: for dimension 4 operators, we have
\be\label{eq:d_coeff}
\left\{\lambda_H = d_H \,g_\star^2\,,\quad \lambda_B = d_B\, g' \frac{g_\star^2}{16\pi^2}\,,\quad  \lambda_B' = d_B' \, g' \frac{g_\star^2}{16\pi^2}\right\},
\ee
where $g'$ is the hyper-charge gauge coupling and all the small $d$'s are O(1) numbers. The same can be done at the level of dimension 6 operators, for which we obtain the expressions summarized in table\,\ref{tab:dim6_h_counting}, where $g_F$ denotes the gauge coupling of the gauge boson $F = G, W, B$ and the $c$'s are O(1) numbers.

\subsection{About the dark photon physical mass}\label{sec:DP_physical_mass}
\begin{table}[tb]
\centering
{\def\arraystretch{1.3}
\begin{tabular}{|c |c|| c |c|| c |c|}
\hline
$\wc_i$ & Power counting & $\wc_i$ & Power counting& $\wc_i$ & Power counting\\
\hline\hline
$\wc_{H}$ & $c_{H}\,g_\star^4 $ & $\wc_{DH}$ & $c_{DH}\,g_\star^2$ & $\wc_{H\psi}$ & $c_{H\psi}\, g_\star^2$ \\
$\wc_{F,\,\tilde F} $ & $g_F^2 \, c_{F,\,\tilde F} \frac{g_\star^2}{16\pi^2}$ & $\wc_{H,2}$ & $c_{H,2} \,g_\star^2 \frac{g_\star^2}{16\pi^2}$ &$\wc_{H,3} $ & $c_{H,3} \,g_\star^2 \frac{g_\star^2}{16\pi^2}$ \\
$\wc_{\psi,\,Y,\,Hc}$ & $c_{\psi,\,Y,\,Hc} \,g_\star^2 $ & $\wc_{\tilde{\psi},\,\tilde Y,\,\tilde Hc} $ & $c_{\tilde{\psi},\,\tilde Y,\,\tilde Hc} \, g_\star^2$ & $\wc_{D2H}$ & $c_{D2H} \, g_\star^2$ \\
$\wc_{T_\psi}$ & $c_{T_\psi}\, g_\star^2$ & $\wc_{T_F}$ & $c_{T_F} \, g_\star^2$ & $\wc_{HB}$ & $c_{HB}\, g'\, g_\star^2 \frac{g_\star^2}{16\pi^2}$ \\
$\wc_{H\psi,2}$ & $c_{H\psi,2}\,g_\star^2$ & $\wc_B$ & $c_B\, g' \frac{g_\star^2}{16\pi^2}$ & $\wc_{HB}'$ & $c_{HB}'\, g'\, g_\star^2 \frac{g_\star^2}{16\pi^2}$ \\
 $\wc_{H\psi,3}$& $c_{H\psi,3}\,g_\star^2$ & $\wc_B'$ & $c_B'\,  g' \frac{g_\star^2}{16\pi^2}$ & $$ & $$\\
\hline
\end{tabular}
}
\caption{List of the normalized effective Wilson coefficients for dimension 6 operators in terms of powers of gauge couplings $g_i$.}
\label{tab:dim6_h_counting}
\end{table} 

The first terms of \eqref{eq:dim4} and \eqref{eq:dim6} renormalize the low-energy free lagrangian of $V_\mu$. Indeed, upon electroweak symmetry breaking they redefine the dark photon mass as 
\be
M^2_{\rm phys}=M^2 \left[1 + d_H \frac{g_\star^2 \vev^2}{M_\star^2} + \frac{c_{H}}{2} \left(\frac{g_\star^2 \vev^2}{M_\star^2} \right)^2\right] .
\ee
With the improved power counting, the physical mass is only partially corrected by higher dimensional operators, since if $M\to 0$ in the above formula the physical mass is itself zero. Also the correction from $c_{H}$ is quadratically suppressed as compared to $d_H$.

There are, however, two more extreme Stueckelberg-like scenarios. First, the case $M\gg g_\star \vev$, where the physical mass originates mostly from dynamics above $M_\star$ in such a way that $M$ corrections to the mass can be neglected. In this limit one could explore a Stuckelberg-like interactions with $|H|^{2n}$ without the suppression $M^{2n}/M_\star^{2n}$. In particular we reach this configuration with the shift $\lambda_H\to \lambda_H^{\rm S} \times M_\star^2/M^2$. The cut-off scale of our EFT becomes $E\gtrsim M/\sqrt{\lambda_H}g_\star$. Second, we have the pure massless Stueckelberg scenario, where $V$ gets its mass entirely from $|H|^2$. This limit can be reached by rescaling $\lambda_H\to \lambda_H^{\rm 0} \times M_\star^2/M^2$ and  $\wc_{H}\to \wc_{H}^{\rm 0} \times M_\star^4/M^4$ and then also $M\to 0$. Such a term has an even lower cut-off than the one discussed in section \ref{sec:stueckelberg}, since in the limit where $M\to 0$ the physical mass does not vanish, but its proportional to $M_{\rm phys}\propto \lambda_H^{\rm 0} g_\star \vev$. Such a limit shows a cut-off in the scattering $hh\to V \bar V$ at around $E\gtrsim \vev$, which is as worse as the one of Higgsless theories. We will briefly come back to this case in section \ref{sec:pheno} where we explore wether this scenario can live in some point of the parameters space tested by experiments. Notice that, upon electro-weak symmetry breaking, the first operator on the second line of \eqref{eq:dim6} renormalizes the wave-function of $V_\mu$. We neglect this correction, since in minimally coupled models it arises at loop-level.


\subsection{Dark Matter abundance}\label{sec:abundance}
Before we go on discussing the DM phenomenology in detail, we would like here to comment on how we expect to reproduce the DM abundance of complex dark photon in our context. We take the point of view - consistent with our EFT approach to the problem - that the DM abundance $\Omega_{\rm DM}$ can be computed only below the scale $M_\star$ of our EFT. Indeed, with our approach we cannot explore the other possibility where contributions to the DM abundance happen also before $M_\star$. However, although we do not rely on them, it is conceivable to imagine that there are mechanisms, both thermal and non-thermal, that contribute to the DM abundance even above $M_\star$.  

We are then left with the chance of producing DM in the early universe only if we consider energy scales, both temperature and Hubble, that are above $M$. There are several distinguished scenarios that can be discussed easily depending on the size of Hubble scale during inflation $H_I$ and the reheating temperature $T_R$. For the practical reasons discussed above we consider $M_\star>T_R$, while the individual size of $T_R$ and $H_I$ compared to the DM mass $M$ is important to identify scenarios of DM production. We here focus on two main cases:

\medskip
\begin{itemize}
\item $T_R<M$: in this case no thermal production of DM is expected and freeze-out from the SM will be negligible. Yet, if $H_I>M$, a massive Stueckelberg vector can be produced gravitationally during inflation, as shown in \cite{Graham:2015rva}. This mechanism only depends on the mass of the vector, and it is at work at the level of the free field, see anyhow \cite{Redi:2022zkt} for a discussion of the impact of self-interactions. The final abundance will depend on the value of the Hubble scale during inflation and on $T_R$\,\cite{Ahmed:2020fhc,Kolb:2020fwh,Bertuzzo:2024fns} and it is possible to find DM masses that reproduce today's abundance of interest for our phenomenological study in the next sections. This mechanism is non-thermal, based on particle production during/after inflation (see also \cite{Kolb:2023ydq} for a review);
\item $T_R>M$: in this case, in addition to the inflationary production, we have also the chance that the DM is kept in thermal equilibrium with the SM bath through the interactions in the EFT. The observed abundance is then thermally produced as the complex dark photon eventually freezes-out. In principle this can be done systematically in the EFT as long as we work in a range $M_\star \gtrsim T_R \gtrsim M$.  The dominant contribution to the DM abundance from freeze-out comes from the EFT operators that contribute to $s$-wave annihilation of complex dark photons into SM final states, during radiation domination. In eq.\,\eqref{eq:EFT}, $s$-wave annihilation comes mainly from the dimension-4 operators Higgs portal and the electric dipole-like ones in eq.~\eqref{eq:dim4}. We focus here briefly on these two, all the others will give smaller annihilation cross-sections and hence larger contributions to the abundance.
We find that, individually, the Higgs portal operator reproduce the DM abundance for a quartic $d_H \lesssim 8\times10^{-5}(M/100\,\GeV)$ while the dipole requires \\$d'_B\lesssim 0.1(M/100\,\GeV)$ with the \textit{naive} power counting. On the other hand, when the \textit{improved} power counting is adopted, the DM abundance is obtained for $d_H\lesssim 8\times10^{-9}(M/100\,\GeV)(M_\star/M)^{2}$ for the Higgs portal and $d'_B\lesssim 10^{-5}(M/100\,\GeV)(M_\star/M)^2$   for the electric-like dipole operator. For values smaller than these, DM is overproduced. As we are going to show in the next section, these values can be larger than the bounds attainable on these Wilson coefficients from direct searches. We conclude that DM thermal production likely leads to overproduction in the region tested by direct detection. In this case, thermal production of DM can be rendered consistent with the observed value requiring dilution of the number density. This however requires extra ingredients beyond the EFT.
\end{itemize}
Our arguments suggest that there is more than a way to reproduce the DM abundance. In the following we assume that today's abundance is always reproduced in our phenomenological studies.


\section{Matching onto low energy EFT}\label{sec:matching}
The EFT so far developed in eq.~\eqref{eq:EFT} cannot be used straightforwardly unless we are just interested in high energy phenomena, as, for example, modifications to invisible branching ratios of $H$ and $Z$ (as discussed in section \ref{subsec:collider}). When we are interested in low energy phenomena like direct detection experiments, we would like to match our eq.~\eqref{eq:EFT} to a non-relativistic theory of nuclei. 

In this section we discuss in detail the procedure to match our EFT onto effective theories valid at even lower energy regimes. While this procedure is completely standard (see \cite{DelNobile:2021wmp} for a technical review), this exercise will allow us to keep track explicitly of all the operators in eq.~\eqref{eq:EFT} by computing their effects on low-energy observables. A summary of the low energy operators generated in our model is shown in two main tables: table \ref{tab:tilde eft} for the EFT of DM and quarks and gluons; table \ref{tab:NRop} for the non-relativistic EFT of DM-nucleons.

\begin{table}[h!tb]
\centering
{\def\arraystretch{1.4}\begin{tabular}{|c |c ||c |c |}
	\hline
       Operator & Expression & Coefficient & Expression \\
       \hline\hline
        $\tilde\Op_1$ & $V_\mu\bar{V}^\mu [m_q\bar{q}q]$ & $\tilde c_1$&$\big[1,\,\frac{M^2}{M_\star^2}\big]\big(-\frac{\lambda_H}{M_h^2}-\frac{\wc_H}{M_\star^2}\frac{\vev^2}{M_h^2}+\frac{\wc_{Hf}}{M_\star^2}\big)$\\
       \hline
        $\tilde\Op_2$ & $V_\mu\bar{V}^\mu[\frac{\alpha_s}{12\pi}G_{\mu\nu}G^{\mu\nu}]$ &$\tilde c_2$&$\big[1,\,\frac{M^2}{M_\star^2}\big]\big(\frac{\lambda_H}{M_h^2}+\frac{\wc_H}{M_\star^2}\frac{\vev^2}{M_h^2}+\frac{\wc_F}{M_\star^2}\big)$\\
        \hline
        $\tilde\Op_3$& $\FS [m_q\bar qq]$ &$\tilde c_3$&$-\frac{\wc_{H,2}}{M_\star^2M_h^2}$\\
        \hline
        $\tilde\Op_4$&$\FS[\frac{\alpha_s}{12\pi}G_{\mu\nu}G^{\mu\nu}]$ & $\tilde c_4$&$-\frac{\wc_{H,2}}{M_\star^2M_h^2}$\\
        \hline
        $\tilde\Op_5$&$\OFS[m_q\bar{q}q]$ & $\tilde c_5$&$-\frac{\wc_{H,3}}{M_\star^2M_h^2}$\\
        \hline
        $\tilde\Op_6$&$\OFS[\frac{\alpha_s}{12\pi}G_{\mu\nu}G^{\mu\nu}]$ & $\tilde c_6$&$-\frac{\wc_{H,3}}{M_\star^2M_h^2}$\\
        \hline
        $\tilde\Op_7$&$J_\mu^V [J_q^\mu]$ & $\tilde c_7$&$\big[1,\,\frac{M^2}{M_\star^2}\big]\frac{1}{M_\star^2}\big(\wc_f+\wc_Y Y_f+\wc_{Hc}\frac{g_Z^2g_v^q\vev^2}{M_Z^2}\big)$\\
        \hline
        $\tilde\Op_8$&$J_\mu^P[ J^\mu_q ]$ & $\tilde c_8$&$\big[1,\,\frac{M^2}{M_\star^2}\big]\frac{g_\star^2}{M_\star^2}\big(\wc_{\tilde f}+\wc_{\tilde Y} Y_f+\wc_{\tilde{Hc}}\frac{g_Z^2g_v^q\vev^2}{M_Z^2}\big)$\\
        \hline
        $\tilde\Op_9$&$iV_{[\mu}\bar{V}_{\nu]}[F^{\mu\nu}]$ & $\tilde c_9$&$\big[1,\,\frac{M^2}{M_\star^2}\big]\frac{c_{\theta_w}}{2}\big(\lambda_B +\wc_{HB}\frac{\vev^2}{M_\star^2}\big)$\\
        \hline
        $\tilde\Op_{10}$&$iV_{[\mu}\bar{V}_{\nu]}[\bar q\sigma^{\mu\nu}q]$ & $\tilde c_{10}$&$\big[1,\,\frac{M^2}{M_\star^2}\big]\frac{\wc_{Hf,2}}{2M_\star^2}m_q$\\
        \hline
        $\tilde\Op_{11}$&$iV_{[\mu}\bar{V}_{\nu]}[\partial_{{\color{white}{A}}}^{[\mu}J_\textup{\tiny{NC}}^{\nu]}]$ & $\tilde c_{11}$&$\big[1,\,\frac{M^2}{M_\star^2}\big]\frac{s_{\theta_w}}{2}\frac{g_Z}{M_Z^2}\big(\lambda_B +\wc_{HB}\frac{\vev^2}{M_\star^2}\big)$ \\
        \hline
        $\tilde\Op_{12}$&$iV^{}_{[\mu\rho}\bar V^\rho_{\,\,\,\nu]}[F^{\mu\nu}]$ & $\tilde c_{12}$&$c_{\theta_w}\frac{\wc_B}{M_\star^2}$\\
        \hline
        $\tilde\Op_{13}$&$iV^{}_{[\mu\rho}\bar V^\rho_{\,\,\,\nu]}[\partial_{{\color{white}{A}}}^{[\mu}J_\textup{\tiny{NC}}^{\nu]}]$ & $\tilde c_{13}$&$s_{\theta_w}\frac{g_Z}{M_Z^2}\frac{\wc_B}{M_\star^2}$\\
        \hline
        $\tilde\Op_{14}$&$V_{(\mu}\bar V_{\nu)}[T_G^{\mu\nu}]$ & $\tilde c_{14}$&$\big[1,\,\frac{M^2}{M_\star^2}\big]\frac{\wc_{TF}}{2M_\star^2}$\\
        \hline
        $\tilde\Op_{15}$&$V_{(\mu}\bar V_{\nu)}[\partial_{{\color{white}{A}}}^{(\mu}J_q^{\nu)}]$ & $\tilde c_{15}$&$\big[1,\,\frac{M^2}{M_\star^2}\big]\frac{\wc_{Tf}}{2M_\star^2}$\\
        \hline
        $\tilde\Op_{16}$&$V_{(\mu}\bar V_{\nu)}[\partial^{(\mu}\partial^{\nu)}m_q\bar q q]$ & $\tilde c_{16}$&$-\big[1,\,\frac{M^2}{M_\star^2}\big]\frac{\wc_{D2H}}{2M_\star^2M_h^2}$\\
        \hline
        $\tilde\Op_{17}$&$V_{(\mu}\bar V_{\nu)}[\partial^{(\mu}\partial^{\nu)}\frac{\alpha_s}{12\pi}G_{\alpha\beta}G^{\alpha\beta}]$ & $\tilde c_{17}$&$-\big[1,\,\frac{M^2}{M_\star^2}\big]\frac{\wc_{D2H}}{2M_\star^2M_h^2}$\\
\hline
        $\tilde\Op_{18}$&$iV_{[\mu}\bar V_{\nu]}\tilde F^{\mu\nu}$ & $\tilde c_{18}$&$\big[1,\,\frac{M^2}{M_\star^2}\big]\frac{c_{\theta_w}}{2}\big(\lambda_B^\prime +\wc_{HB}^\prime\frac{\vev^2}{M_\star^2}\big)$\\
\hline
$\tilde\Op_{19}$&$\varepsilon^{\mu\nu\rho\sigma}iV_{[\mu}\bar V_{\nu]}[\bar q \sigma_{\rho\sigma}q]$ & $\tilde c_{19}$&$-\big[1,\,\frac{M^2}{M_\star^2}\big]\frac{\wc_{H\psi,3}}{2M_\star^2}m_q$\\
\hline
$\tilde\Op_{20}$&$\varepsilon^{\mu\nu\rho\sigma}iV_{[\mu}\bar V_{\nu]}[\partial_{[\mu}J_{{\text{\tiny{NC}}}\,\sigma]}]$ & $\tilde c_{20}$&$\big[1,\,\frac{M^2}{M_\star^2}\big]\frac{s_{\theta_w}}{2}\frac{g_Z}{M_Z^2}\big(\lambda_B^\prime +\wc_{HB}^\prime\frac{\vev^2}{M_\star^2}\big)$\\
        \hline
        $\tilde\Op_{21}$&$iV^{}_{[\mu\rho}\bar V^\rho_{\,\,\,\nu]}[\tilde F^{\mu\nu}]$ & $\tilde c_{21}$&$c_{\theta_w}\frac{\wc_B^\prime}{M_\star^2}$\\
        \hline
        $\tilde\Op_{22}$&$\varepsilon^{\mu\nu\rho\sigma}iV^{}_{[\mu\alpha}\bar V^\alpha_{\,\,\,\nu]}[\partial_{[\rho}J_{\text{\tiny{NC}}\,\sigma]}]$ & $\tilde c_{22}$&$s_{\theta_w}\frac{g_Z}{M_Z^2}\frac{\wc_B^\prime}{M_\star^2}$\\
        \hline
	\end{tabular}}
\caption{Effective operators involving quarks and gluons, and corresponding Wilson coefficients, obtained matching the EFT presented in section \ref{sec:structure_EFT} to the low energy EFT of eq.\,\eqref{eq:EFT_below_MZ} at the weak scale. We set the Wilson coefficients $\wc_{DH},\,\wc_{\tilde F}=0$ since they will not play any role in the rest of the work. In the second column we indicate in square brackets the two alternative choices for the power counting, i.e. $1$ for the \textit{naive} power counting and $M^2/M_\star^2$ for the $improved$ power counting.}
 \label{tab:tilde eft}
\end{table}
\subsection{Matching at the Electro-Weak Scale}\label{sec:tildeeft}
According to our fundamental assumptions, our EFT is defined at the UV scale $M_\star\gg M_\textup{\tiny{EW}}$ at which any heavy dark degree of freedom is integrated out. Moving to lower energies, we need to perform a first matching procedure at the electroweak scale, when the $Z$, $W$, $h$ bosons and the heavy top-quark must be integrated out. The effective Lagrangian valid at energies $E\ll M_\textup{\tiny{EW}}$ is generally given by:
\be\label{eq:EFT_below_MZ}
\Lag_\textup{\tiny{EFT}}\big|_\textup{$E\ll M_\textup{\tiny{EW}}$}\supset\sum_{i=1}^{22} \tilde c_i\,\tilde{\Op}_i ,
\ee
with the operators $\tilde{\Op}_i$ and the corresponding Wilson coefficients $\tilde{c}_i$ listed in table \ref{tab:tilde eft}. In the table, we write the SM structures to be further evaluated when even lower energies are considered (see next section) in square brackets, and we also show in square brackets the two possible power counting choices according to what stated in table \ref{table:power-counting}. 

In order to set the notation, $\theta_w$ is the weak angle, the quark current $J_q^\mu$ that appears in $\tilde{\Op}_{7, 8, 11, 13, 15}$ is the vector current $J_q^\mu = \bar{q} \gamma^\mu q$ computed with light quarks, while $J_\textup{\tiny{NC}}^\mu$ is the neutral current to which the $Z$ boson couples,
\be
J_\textup{\tiny{NC}}^\mu = \left(\frac{1}{2} - \frac{4}{3} s_{\theta_w}\right) \,\bar{u}\gamma^\mu u +  \left(-\frac{1}{2} +\frac{2}{3} s_{\theta_w}\right)\left(\bar{d} \gamma^\mu d + \bar{s} \gamma^\mu s\right).
\ee
It appears in the $\tilde\Op_{11,13,20,22}$ operators because they are obtained integrating out the $Z$ boson in the s-channel (as well as in the contribution from the Higgs current in $\tilde\Op_{7,8}$ but with a different structure, so we dropped the notation momentarily there). We notice that from each coupling of the complex dark photon to the hypercharge, we obtain two contributions below the EW symmetry breaking, among which the one due to the Z-coupling is evidently suppressed by an extra $1/M_Z^2$ factor. On the other hand, the operators $\tilde{O}_{9,12,18,21}$ that contain a single photon field strength will be particularly important for direct detection, since they contribute to the cross section via a photon exchange in the t-channel and are thus {\it enhanced}, rather than suppressed, by the small momentum exchanged in the reaction. On top of that, we will only show bounds coming from the $\gamma$-coupling when discussing direct detection phenomenology in section \ref{subsec:DD}.
Finally, $T_G^{\mu\nu}$ denotes the energy-momentum tensor of gluons. We do not consider operators that contain leptons and more than one photon because they are not relevant (at leading order) for the phenomenology we discuss in this work.

\subsection{Single Nucleon EFT}\label{sec:nucleonEFT}
We now take the EFT defined by eq.\,\eqref{eq:EFT_below_MZ} and move to even lower energies. The next relevant threshold we encounter is around the GeV, i.e. around the QCD confinement scale, when we must match to a single-nucleon relativistic EFT. In momentum space, the Lagrangian can be written in terms of the nucleon field $N$ as 
\be
\begin{split}
\Lag_N=& c_SV_{\mu}\bar{V}^\mu\bar{N}N+c_{FS}\FS\bar NN+c_{PS}\OFS\bar NN+c_VJ_\mu^VJ_N^\mu+c_{PV}J_\mu^PJ_N^\mu+c_{M}V_{[\mu}\bar{V}_{\nu]}q_{{\color{white}{A}}}^{[\mu}J_N^{\nu]}\\
&+c_{AT}V_{[\mu}\bar{V}_{\nu]}\bar{N}\sigma^{\mu\nu}N+c_M^\prime V^{}_{[\mu\rho}\bar{V}^\rho_{\,\,\,\nu]}q_{}^{[\mu}J_N^{\nu]}+c_{ST}V_{(\mu}\bar{V}_{\nu)}q_{}^{(\mu}J_N^{\nu)}+c_{2d}V_{(\mu}\bar{V}_{\nu)}q^{(\mu}q^{\nu)}\bar NN\\
&+c_E\,\varepsilon^{\mu\nu\rho\sigma}V_{[\mu}\bar{V}_{\nu]}q_{[\rho}J_{N\,\sigma]}+c_{AT}^\prime\,\varepsilon^{\mu\nu\rho\sigma}V_{[\mu}\bar{V}_{\nu]}\bar{N}\sigma_{\rho\sigma}N+c_E^\prime\,\varepsilon^{\mu\nu\rho\sigma} V^{}_{[\mu\alpha}\bar{V}^\alpha_{\,\,\,\nu]}q_{[\rho}J_{N\,\sigma]}\\
\end{split}
\label{eq:nucleon lag}
\ee
where $q$ represents the exchanged 4-momentum and the vector nucleon current is $J_N^\mu = \bar{N} \gamma^\mu N$ . The set of effective operators appearing in eq.\,\eqref{eq:nucleon lag} is obtained by evaluating the operator structures in square brackets containing quarks and gluons that appear in table \ref{tab:tilde eft} using the matrix elements listed in\,\cite{Nobile:2013aa,DelNobile:2021wmp}.
The dimensionful Wilson coefficients appearing in eq.\,\eqref{eq:nucleon lag} can then be written in terms of the tilded Wilson coefficients of table \ref{tab:tilde eft} as
\begin{align}
\begin{aligned}\label{eq:nucleon eft coeff}
c_S & =m_N\left[\left(\sum_{q}f_{Tq}^N \right) \tilde c_1 +\frac{2}{27}\left(\sum_{q}f_{Tq}^N-1\right)\tilde c_2\right]\,, & ~~~  c_V & =\sum_{q}F_1^{q,N}(0)\,\tilde c_7\,,\\
c_{FS} & =m_N\left[\left(\sum_{q}f_{Tq}^N \right) \tilde c_3+\frac{2}{27}\left(\sum_{q}f_{Tq}^N-1\right)\tilde c_4\right]\,, & ~~~  c_{PV} & =\sum_{q}F_1^{q,N}(0)\,\tilde c_8\,,\\
c_{PS} & =m_N\left[\left(\sum_{q}f_{Tq}^N\right) \tilde c_5+\frac{2}{27}\left(\sum_{q}f_{Tq}^N-1\right)\tilde c_6\right]\,, & ~~~  c_{AT} & =\sum_{q}F_{T,0}^{q,N}(0)\,\tilde c_{10}\,,\\
c_{2d} & =m_N\left[\left(\sum_{q}f_{Tq}^N\right) \tilde c_{16}+\frac{2}{27}\left(\sum_{q}f_{Tq}^N-1\right)\tilde c_{17}\right], & ~~~  c_{ST} & =\sum_{q}F_1^{q,N}(0)\,\tilde c_{15}\,, \\
c_M & =\frac{Q_N}{q^2} \tilde c_9 +F_1^{\textup{\tiny{NC}}}(0)\, \tilde c_{11}\,,& ~~~ 
c_M^\prime & =\frac{Q_N}{q^2} \tilde c_{12}+F_1^{\textup{\tiny{NC}}}(0)\,\tilde c_{13}\,,\\
c_E & =\frac{Q_N}{q^2} \tilde c_{18} +F_1^{\textup{\tiny{NC}}}(0)\, \tilde c_{20}\,,& ~~~ 
c_E^\prime & =\frac{Q_N}{q^2} \tilde c_{21}+F_1^{\textup{\tiny{NC}}}(0)\,\tilde c_{22}\,,\\
c_{AT}^\prime&=\sum_{q}F_{T,0}^{q,N}(0)\,\tilde c_{19}\,\,,&
\end{aligned}
\end{align}
where we used the standard notation for the form factors that appear at the nucleon level (see\,\cite{Nobile:2013aa,DelNobile:2021wmp}). 
The effective coefficients generated by UV interactions to the hypercharge are here labelled according to the type of interaction they resemble, \ie $c^{(\prime)}_M$ contributes to the complex dark photon magnetic dipole while $c^{(\prime)}_E$ is the triggered electric dipole-like coupling.

\subsection{Non-relativistic Effective Field Theory}\label{sec:NREFT}
\begin{table}[tb]
\centering
\begin{tabular}{| l|c |c || l | l|c | l|c}
\hline
$\Op_i^\textup{\tiny{NR}}$ & Structure & $\wc_i$ & $\Op_i^\textup{\tiny{NR}}$ & Structure & $\wc_i$ \\
\hline\hline
$\cellcolor{light!20}\mathcal{O}_1^\textup{\tiny{NR}}$ & \cellcolor{light!20}$1$ & $\cellcolor{light!20}\text{\small{$\lambda_H,\,\wc_{H,2},\,\wc_F,\,\wc_\psi$}}$ & $\mathcal{O}_{13}^\textup{\tiny{NR}}$ & $i(\SV\cdot \vec v_\perp)(\vec q\cdot \vec s_N)/m_N$ & $$ \\
\cellcolor{light!20}$\mathcal{O}_2^\textup{\tiny{NR}}$ & \cellcolor{light!20}$v_\perp^2$ & \cellcolor{light!20} \small{not generated} & $\mathcal{O}_{14}^\textup{\tiny{NR}}$& $i(\vec s_N\cdot \vec v_\perp)(\vec q\cdot \SV)/m_N$ & $$ \\
$\mathcal{O}_3^\textup{\tiny{NR}}$ & $ i \vec s_N \cdot (\vec q \times \vec v_\perp)/m_N$ & $$ &$\mathcal{O}_{15}^\textup{\tiny{NR}}$ & $(\SV \cdot \vec q)[(\vec s_N \times \vec v_\perp)\cdot \vec q]/m_N^2$ & $$ \\
$\mathcal{O}_4^\textup{\tiny{NR}}$ & $\SV \cdot \vec s_N$ & $$ & $\mathcal{O}_{16}^\textup{\tiny{NR}}$ & $[\SV \cdot (\vec q \times \vec v_\perp)](\vec q\cdot \vec s_N)/m_N^2$ & $$ \\
\cellcolor{light!20}$\mathcal{O}_5^\textup{\tiny{NR}}$  & \cellcolor{light!20}$ i \SV \cdot (\vec q \times \vec v_\perp)/m_N$ & \cellcolor{light!20}$ \text{\small{$\lambda_B,\,\wc_B$}}$ & $\cellcolor{light!20}\mathcal{O}_{17}^\textup{\tiny{NR}}$ & \cellcolor{light!20}$i q^i \mathcal{S}_{ij} v_\perp^j/m_N$ & \cellcolor{light!20}$\text{\small{$\wc_{T\psi}$}}$ \\
$\mathcal{O}_6^\textup{\tiny{NR}}$ & $(\SV \cdot \vec q)(\vec s_N \cdot \vec q)/m_N^2$ & $$ & $\mathcal{O}_{18}^\textup{\tiny{NR}}$ & $i q^i \mathcal{S}_{ij} s_N^j/m_N$ & $$ \\
$\mathcal{O}_7^\textup{\tiny{NR}}$ & $\vec s_N \cdot \vec v_\perp$ & $$ & \cellcolor{light!20}$\mathcal{O}_{19}^\textup{\tiny{NR}}$ &\cellcolor{light!20} $q^iq^j \mathcal{S}_{ij}/m_N^2$ & \cellcolor{light!20}$\text{\small{$\wc_{H\psi,2},\,\wc_{D2H}$}}$\\
\cellcolor{light!20}$\mathcal{O}_8^\textup{\tiny{NR}}$ & \cellcolor{light!20}$\SV \cdot \vec v_\perp$ & \cellcolor{light!20}$\text{\small{$\wc_{\tilde \psi}$}}$ & $\mathcal{O}_{20}^\textup{\tiny{NR}}$ & $[(\vec s_N\times \vec q)]^i q^j\mathcal{S}_{ij}/m_N^2$ & $$ \\
$\mathcal{O}_9^\textup{\tiny{NR}}$ & $i\SV \cdot (\vec s_N \times \vec q)/m_N$ & $$ & $\mathcal{O}_{21}^\textup{\tiny{NR}}$ & $v_\perp^i s_N^j \mathcal{S}_{ij}$ & $$ \\
$\mathcal{O}_{10}^\textup{\tiny{NR}}$ & $i(\vec q\cdot \vec s_N)/m_N$ & $$ & $\mathcal{O}_{22}^\textup{\tiny{NR}}$ & $i[(\vec q \times \vec v_\perp)]^i s_N^j\mathcal{S}_{ij}/m_N$ & $$ \\
 \cellcolor{light!20}$\mathcal{O}_{11}^\textup{\tiny{NR}}$ & \cellcolor{light!20}$i(\vec q\cdot \SV)/m_N$ & \cellcolor{light!20}$\text{\small{$\wc_{H,3},\,\lambda_B^\prime,\,\wc_B^\prime,\,\wc_{H\psi,3}$}}$ & $\Op_{23}^\textup{\tiny{NR}}$ &  $i[(\vec s_N \times \vec v_\perp)]^i q^j \mathcal{S}_{ij}/m_N$ & $$\\
 $\mathcal{O}_{12}^\textup{\tiny{NR}}$ & $\SV \cdot (\vec s_N \times \vec v_\perp)$ & $$ & $\Op_{24}^\textup{\tiny{NR}}$ &  $i[(\vec s_N \times \vec q)]^i v_\perp^j \mathcal{S}_{ij}/m_N$ & $$\\
& & &\cellcolor{light!20}$\Op_{25}^\textup{\tiny{NR}}$ &\cellcolor{light!20}$v_\perp^i v_\perp^j \mathcal{S}_{ij}$ & \cellcolor{light!20}\small{not generated}\\
\hline
\end{tabular}
\caption{Basis of non-relativistic operators for direct detection of complex spin-1 DM. The first column shows the usual non redundant basis constructed with a single structure. Operators $\Op_i$ exclusive for spin-1 DM start from $i\geq 17$. Spin-independent operators have been highlighted in green and we provide the leading contribution from UV Wilson coefficient that generate each specific interaction, with the complete matching given in eq. \eqref{eq:matchingNR}.}
\label{tab:NRop}
\end{table}

Finally, the relativistic EFT defined in eq.\,\eqref{eq:nucleon lag} should be reduced to a non-relativistic EFT (NREFT), that can be used to compute nuclear response functions needed for the purpose of direct detection. Such a NREFT can be written in terms of the following independent Galilean invariants:
\be\label{eq:NREFT_structures}
\left\{\, \mathcal I_{V,N},\,\, \vec q = \vec p -\vec p\,' = \vec k -\vec k\,'\,,\,\,  \vec P = \vec p + \vec p\,'\,,\,\, \vec v_\perp = \frac{\vec P}{2M}-\frac{\vec P_N}{2m_N}\,,\,\, \vec P_N =\vec k +\vec k\,'\,, \,\, \vec S_V\,,\,\, \vec s_N\,, \,\, \mathcal{S}\, \right\}\,,
\ee
where $\mathcal I_{V,N}$ is the identity operator acting in the dark photon (V) or nucleon (N) space, $\vec{p}$ and $\vec{p}\,'$ are the initial and final DM momenta, $\vec{k}$ and $\vec{k}'$ are the initial and final nuclear momenta, $\SV$ and $\mathcal{S}$ are the complex dark photon spin operators (we defer to appendix \ref{app:polarizations} for the definition of such operators for spin-1 DM), $\vec{s}_N$ represents the nucleon spin and $\vec v_\perp$ is the so called transverse velocity.

The NREFT Lagrangian is constructed out of Galilean and rotational invariant combinations of the structures presented in eq.\,\eqref{eq:NREFT_structures} and explicitly reads 
\be\label{eq:NREFT}
\mathscr{L}_{\rm NR}=\sum_{i=1}^{25}\sum_N c_i^N \Op_i^\textup{\tiny{NR}}\,\,,
\ee
This ``Lagrangian'' must be interpreted as the matrix element of the direct detection scattering process, as is clear from its dimensions, computed for the a set of hermitian operators that were classified in \cite{Fan:2010gt,Fitzpatrick:2012ix} and augmented in \cite{Dent:2015zpa,Catena:2019hzw,Gondolo:2020wge} to take into account the additional structures that appear for spin-1 DM. We provide the complete set of invariant structures in table \ref{tab:NRop}. Since our focus in section \ref{subsec:DD} will be exclusively on spin-independent cross-sections, in the table we highlight in green the operators that are relevant for our purpose and we point to the relative leading contribution in terms of the coefficients for the high energy effective theory presented in section \ref{sec:structure_EFT}.

We match the single nucleon EFT of eq.\,\eqref{eq:nucleon lag} onto eq.\,\eqref{eq:NREFT} and obtain the following results for the Wilson coefficients generating spin-independent interactions:
\begin{align}\label{eq:matchingNR}
\begin{aligned}
c_1^N &= -2m_N\left(c_S +2M^2c_{FS}+2Mc_V\right) &&\leftarrow&&\left\{\lambda_H,\wc_H,\wc_{H\psi},\wc_F,\wc_{H,2},\wc_{\psi,Y,Hc}\right\}\,,\\
c_2^N &= \textrm{Not generated} &&\leftarrow&&\left\{\right\}\,,\\
c_5^N &= 4m_N^2\left(-c_M +M^2c_M^\prime\right)&&\leftarrow&&\left\{\lambda_B,\,\wc_{HB},\,\wc_B\right\}\,,\\
c_8^N &= -4Mm_Nc_{PV} &&\leftarrow&&\left\{\wc_{\tilde \psi,\tilde Y\tilde Hc}\right\}\,,\\
c_{11}^N &= -2m_N\left(4Mm_Nc_{PS}+c_{AT}^\prime+2m_Nc_E+2M^2m_Nc_E^\prime\right) &&\leftarrow&&\left\{\wc_{H,3},\,\wc_{H\psi,3},\,\lambda_B^\prime,\,\wc_{HB}^\prime,\,\wc_B^\prime\right\}\,,\\
c_{17}^N &=-8m_N^2c_{ST} &&\leftarrow&&\left\{\wc_{T\psi}\right\}\,,\\
c_{19}^N &= 2m_N^2\left(4m_Nc_{2D}+\frac{c_{AT}}{M}\right) &&\leftarrow&&\left\{\wc_{D2H},\,\wc_{H\psi,2}\right\}\,,\\
c_{25}^N &= \textrm{Not generated} &&\leftarrow&&\left\{\right\} \,.\\
\end{aligned}
\end{align}
For each entry, we also list the high-energy Wilson coefficients that contribute to the low energy coefficients $c_i^N$. 
We observe that $c_5^N$ is enhanced by a factor $1/\vec{q}\,^2$ contained in $c_{M}^{(\prime)}$ and the same applies to the $c_E^{(\prime)}$ contribution to the NR Lagrangian coefficient $c_{11}^N$. This factor comes from the photon exchange and is generated by the magnetic $(\tilde{\mathcal{O}}_9,\,\tilde{\mathcal{O}}_{12})$ and electric ($\tilde{\mathcal{O}}_{18},\,\tilde{\mathcal{O}}_{21}$) ``dipole-like'' interactions appearing in table \ref{tab:tilde eft}, respectively.
We stress that we keep only the leading contributions generated by each of the operators in eq.\,\eqref{eq:nucleon lag} in a $q^n$ expansion, neglecting for instance higher order terms like $c_S\Op_{19}^\textup{\tiny{NR}}\sim O(q^2)$ or $c_M q^2\Op_1^\textup{\tiny{NR}}$ which does not show the $1/q^2$ enhancement for dipole-like interactions we just mentioned.

\section{Phenomenological Bounds}\label{sec:pheno}
The main focus of this section are phenomenological constraints on our EFT. In this section we derive bounds on the Wilson coefficients with particular emphasis on the improved power counting.
For this very same reason, prior to the discussion of direct detection, we also consider collider bounds coming from invisible decays of the Higgs and $Z$ bosons. This will serve us as a proof of the importance of this approach in deriving sensible limits, as we will show how an incorrect treatment of the cut-off of massive Stueckelberg would lead to spurious very strong bounds.

In this section, in order to derive bounds from direct detection we will assume that for each point in the parameter space we are able to reproduce the DM abundance (see section \ref{sec:abundance} for a discussion).

Irrespectively of the Stueckelberg nature of the complex dark photon, our limits should only be applied when $M\ll M_\star$, while we are free to vary $g_\star$ from a weak to a moderate strong coupling regime, without invalidating the EFT approach.
In what follows, we will always fix $g_\star$ to two specific values (to be discussed below) and show bounds obtained fixing the value of the Wilson coefficients $d_i=1$, $c_i = 1$, leaving $M_\star$ free to vary, so that exclusion curves will thus be read as lower bounds on the size of $M_\star$.  Clearly, the validity of the EFT requires that the new physics scale, \ie the mass of the heavy mediator $M_\star$, must be much larger than the lighter complex dark photon degree of freedom. We here define for convenience the ratio 
\be
R\equiv\frac{M_\star}{M}\,,
\ee
 in such a way that the EFT description is valid as long as $R \gg 1$. We will comment later on the effect of changing the value of $g_\star$. Each bound will be shown for both power counting schemes (naive and improved) discussed in sections.\,\ref{sec:stueckelberg} and\,\ref{sec:power_counting}.\\

We start our discussion in sec.\,\ref{subsec:collider} by discussing the limits coming from the invisible decays of the Higgs and $Z$ bosons, and then in sec.\,\ref{subsec:DD} we will instead compute bounds coming from direct detection experiments, following the same prescription for the size of the Wilson coefficients.

\subsection{Collider Constraints: Higgs and $Z$ invisible decays }\label{subsec:collider}

\begin{table}[t]
\centering
{\def\arraystretch{1.8}
\begin{tabular}{|l|l|l|}
\hline
\multicolumn{1}{|c|}{  Operator} & \multicolumn{1}{c|}{  Coefficient $\alpha_i$}  & \multicolumn{1}{c|}{  Kinematic factor $\eta_i$} \\
\hline\hline
 $\Op_S \left(|H|^2,  |H|^4\right)$ & $\alpha_1=\left[1,\frac{M^2}{M_\star ^2}\right]\,g_\star^2\,\vev\,\big(d_H+ c_H \frac{g_\star^2\,\vev^2}{M_\star ^2}\big)$ &$\eta_1=\frac{\left(M_h^2-2 M^2\right)^2}{4 M^4}+2$\\
$\Op_{F^2}|H|^2$& $\alpha_2=\frac{g_\star^4}{16 \pi ^2}\,c_{H,2} \,\frac{\vev}{M_\star ^2}$ & $\eta_2=2M^4\left(6-4\frac{M_h^2}{M^2}+\frac{M_h^4}{M^4}\right)$\\
$\Op_P|H|^2$ & $\alpha_3=\frac{g_\star^4}{16 \pi ^2}\,c_{H,3}\,\frac{\vev}{M_\star ^2}$ & $\eta_3=8M_h^4\left(1-4\frac{M^2}{M_h^2}\right)$\\
$\Op_{\mu\nu}^S\partial^{(\mu}\partial^{\nu)}|H|^2$ & $\alpha_4=\left[1,\frac{M^2}{M_\star ^2}\right]\,c_{D2H}\,\frac{g_\star^2\,\vev}{2\,M_\star ^2}$ & $\eta_4= \frac{M_h^4(M_h^2-4 M^2)^2}{4 M^4}$\\
\hline
\end{tabular}
}
\caption{List of the parameters appearing in the invisible decay width of the Higgs boson, see eq.\,\eqref{H inv BR}. For each set of parameters we also present the operator(s) that generate them. We also show in square brackets the two alternative solutions for the naive and improved power counting when a choice is needed.}
\label{tab:hdecay}

\centering
{\def\arraystretch{1.6}\begin{tabular}{|l|l|l|}
\hline
\multicolumn{1}{|c|}{  Operator} & \multicolumn{1}{c|}{  Coefficient $\beta_i$}  & \multicolumn{1}{c|}{  Kinematic factor $\rho_i$} \\
\hline\hline
 $J_\mu^VJ_H^\mu$ & $\beta_1=\left[1,\frac{M^2}{M_\star^2}\right] \sqrt{g^2+g'^2}\,c_{Hc}\,\frac{g_\star^2\,\vev^2}{2\,M_\star^2}$ & $\rho_1=\frac{M_Z^6-8 M^2 M_Z^4+28 M^4 M_Z^2-48 M^6}{12 M^4}$\\
$J_\mu^PJ_H^\mu$& $\beta_2=\left[1,\frac{M^2}{M_\star^2}\right]\sqrt{g^2+g'^2}\,c_{\tilde{H}c}\frac{g_\star^2\,\vev^2}{2\,M_\star^2}$ & $\rho_2=\frac{(M_Z^2-4 M^2)^2}{3 M^2}$\\
$\Op^A_{\mu\nu}B^{\mu\nu} \left(1,|H|^2\right)$ & $\beta_3=\big[1,\,\frac{M^2}{M_\star^2}\big]\frac{g_\star^2\, g'}{16 \pi ^2}\,s_{\theta_w} \big(d_B+c_{HB}\frac{g_\star^2\,\vev^2}{2\,M_\star^2}\big)$ & $\rho_3=\frac{M_Z^2}{12}\left(\frac{M_Z^4}{M^4}-16\right)$\\
$\Op_{\mu\nu}^TB^{\mu\nu}$ & $\beta_4=\frac{g_\star^2\,g'}{16 \pi ^2}s_{\theta_w}\frac{c_B}{M_\star^2}$ & $ \rho_4=\frac{M_Z^4}{6} \left(M_Z^2-8 \frac{M^4}{M_Z^2}-2 M^2\right)$\\
$\Op^A_{\mu\nu}\tilde B^{\mu\nu} \left(1,|H|^2\right)$ & $\beta_5=\left[1,\frac{M^2}{M_\star^2}\right]\frac{g_\star^2\, g'}{16 \pi ^2}\,s_{\theta_w} \big(d_B^\prime+c_{HB^\prime}\frac{g_\star^2\,\vev^2}{2\,M_\star^2}\big)$ & $\rho_5=\frac{1}{3} \left(\frac{M_Z^4}{M^2}+2M_Z^2\right)$ \\
$\Op_{\mu\nu}^T\tilde B^{\mu\nu}$ & $\beta_6=\frac{g_\star^2\, g'}{16 \pi ^2}s_{\theta_w}\frac{c_B^\prime}{M_\star^2}$ & $\rho_6=\frac 23 M_Z^4\left(1-2\frac{M^2}{M_Z^2}+4\frac{M^4}{M_Z^4}\right)$\\
\hline
\end{tabular}
}
\caption{As in table \ref{tab:hdecay} but for the $Z\to V\bar{V}$ decay, see eq.\,\eqref{Z inv BR}. The couplings $g$ and $g'$ are the usual $SU(2)_L \times U(1)_Y$ gauge couplings, while $s_{\theta_w}$ is the sine of the weak angle.}
\label{tab:Zdecay}
\end{table}
In this section we consider the effect of DM in the invisible branching ratios of Higgs and $Z$ boson. As such, we are applying our EFT in an energy range where $M\ll E$ and our improved power counting has to be used. In order to show its effect on the limits, we present a comparison between what happens with and without such power counting fixing $g_\star=1$.

When the Higgs and $Z$ bosons are heavier than twice the DM mass, all dimension 4 operators in eq. \eqref{eq:dim4} and some of the effective operators in eq. \eqref{eq:dim6} may induce their decay into a particle-antiparticle pair of DM. We can then use the measured value of the decays into invisible states to put bounds on the parameter space of the EFT we are considering. We write the two relevant decay widths as
\begin{subequations}\label{eq:invisible_decay_width}
\begin{eqnarray}
\Gamma_{h\to V\bar V}&=\frac{1}{16\pi M_h} \left(\sum_i |\alpha_i|^2\eta_i \right)\sqrt{1-\frac{4 M^2}{M_h^2}} \label{H inv BR}\,\,,\\
\Gamma_{Z\to V\bar V}&=\frac{1}{16\pi M_Z} \left(\sum_i |\beta_i|^2\rho_i\right) \sqrt{1-\frac{4 M^2}{M_Z^2}} \label{Z inv BR}\,\,,
\end{eqnarray}
\end{subequations}
where $M_h$ and $M_Z$ are the Higgs and $Z$ bosons masses, respectively. The values of the coefficients $\alpha_i$, $\beta_i$, and the kinematic structures $\eta_i$ and $\rho_i$ are collected in tables \ref{tab:hdecay} and \ref{tab:Zdecay}, together with the operators that generate the decays. For the coefficients $\alpha_i$ and $\beta_i$ we show in brackets the two alternative choices associated with the naive and improved power counting solutions according to table \ref{table:power-counting}.

We have written the decay widths in this form to highlight two effects. First, the coefficients $\alpha_i$ and $\beta_i$ are just avatars for the Wilson coefficients of eq. \eqref{eq:EFT}, and as such they are linear in them, as can be seen from tables \ref{tab:hdecay} and \ref{tab:Zdecay}. Second, the coefficients $\rho_i$ and $\eta_i$ are instead kinematic factors, that only depend on $M$, and they differ from case to case. In particular, by looking at table \ref{tab:hdecay} and table \ref{tab:Zdecay}, we see that they might display apparent divergences as $M\ll M_{h,Z}$. This makes self-evident the situation that might arise by extrapolating bounds fixing $\alpha_i,\beta_i$ and letting $M$ drop arbitrarily.

Assuming that the decay into the $V\bar{V}$ pair saturates the $h$ and $Z$ invisible widths, we can set bounds using the experimental values $BR_{h \to {\rm inv}} < 13\%$\,\cite{Workman:2022ynf} and $\Gamma_{Z \to V \bar V} < \delta\Gamma(Z \to {\rm inv})$, where $\delta\Gamma_{Z \to {\rm inv}} = 1.5$ MeV is the error on the $Z$ invisible width\,\cite{Workman:2022ynf}. 

\begin{figure}[p]
\begin{center}
\includegraphics[width=0.45\textwidth]{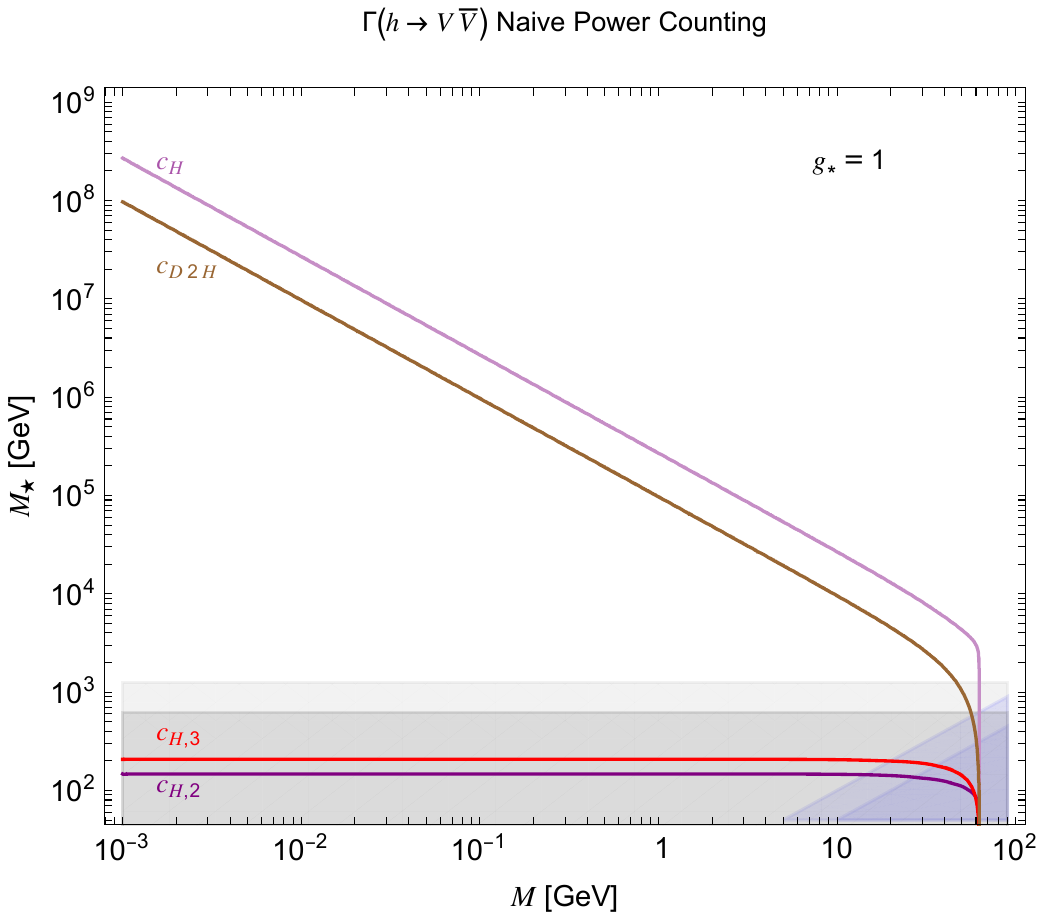}
\includegraphics[width=0.45\textwidth]{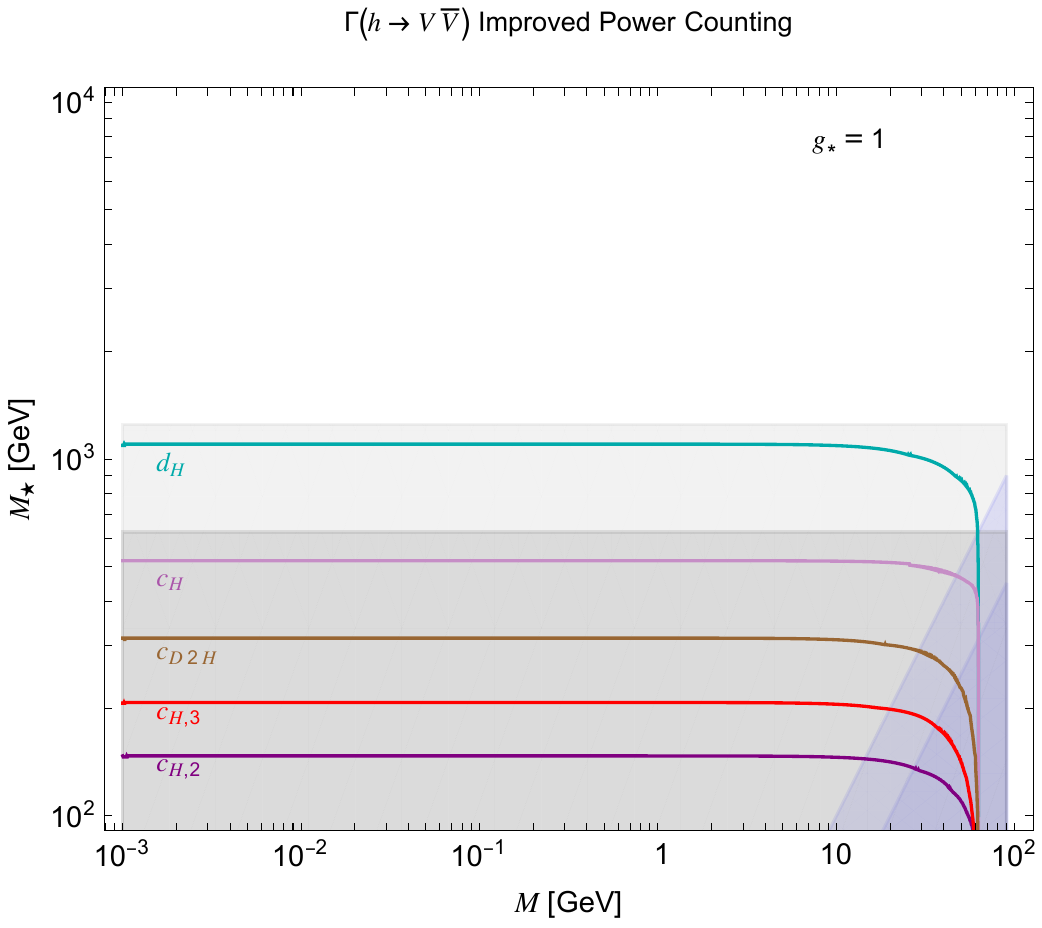}\\
\includegraphics[width=0.45\textwidth]{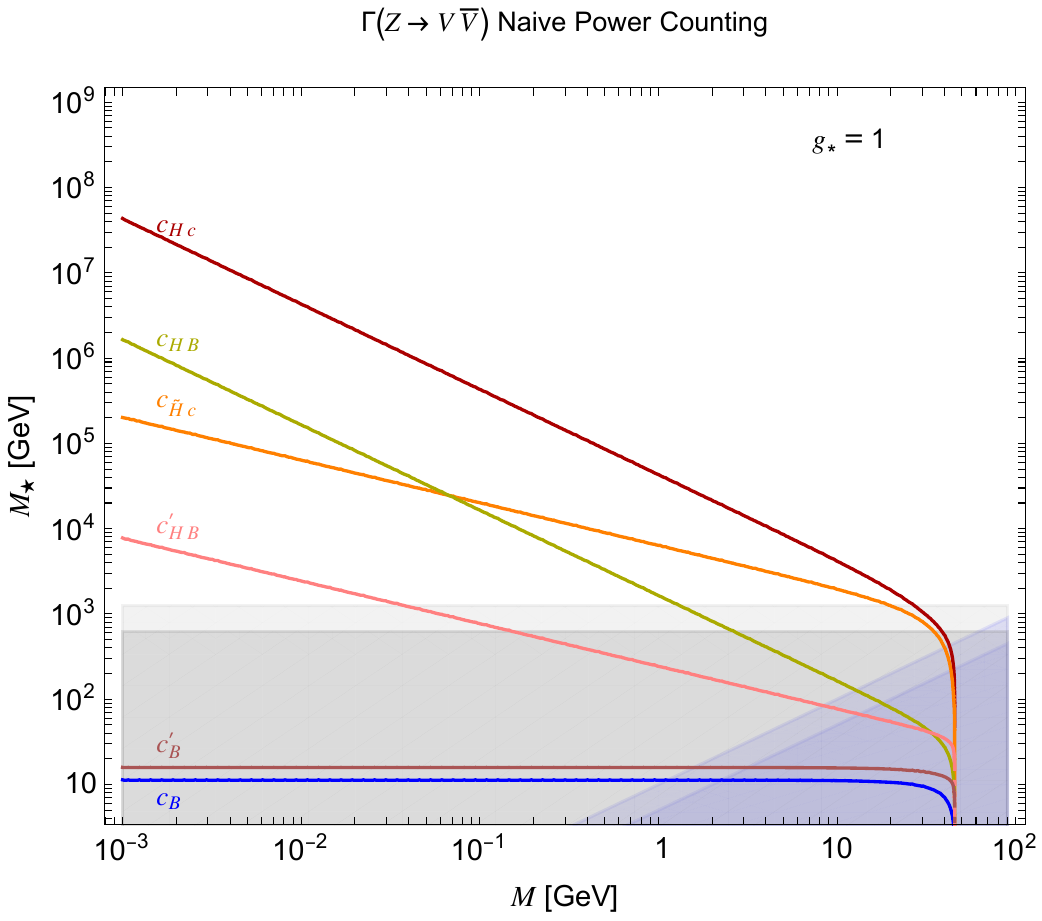}
\includegraphics[width=0.45\textwidth]{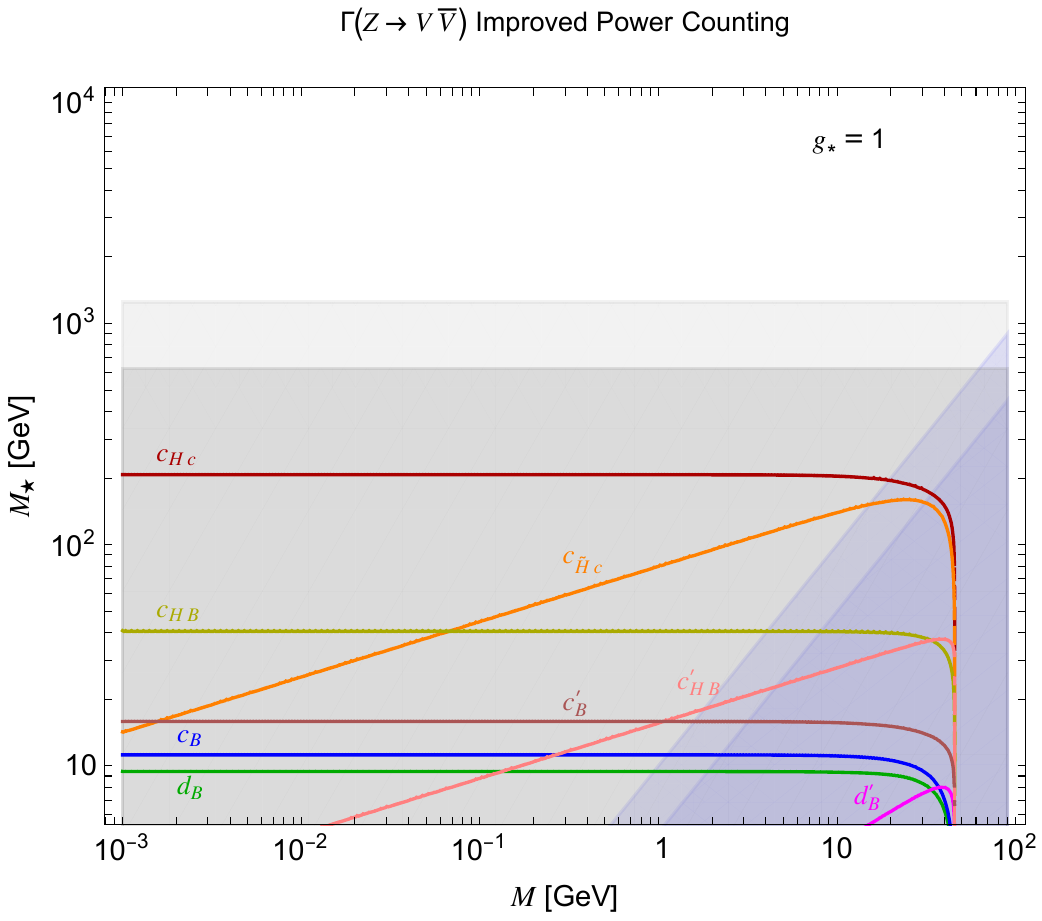}\\
\caption{Regions excluded by the bound on the invisible decays of the Higgs boson (upper panels) or $Z$ boson (lower panels). The regions excluded are those below the colored lines. In the left (right) panel we show the limits obtained using the naive (improved) power counting. All the limits are obtained fixing $g_\star = 1$, $d_i=1$ and $c_i = 1$. The grey horizontal regions show when $M_\star < 5 M_{h,Z}$ or $M_\star < 10 M_{h,Z}$, while the oblique lines show the regions in which $R<5$ (lightblue) or $R<10$ (lighter blue). }
\label{fig:BRL}
\end{center}
\end{figure}

Our results are presented in Fig.~\ref{fig:BRL} where we show the constraints for each operator in the plane $(M,M_\star)$. In such figure all the dimensionless coefficients are set to one, allowing us to display all the constraints in the same plot. In the left panels of figure \ref{fig:BRL} we show the limits on $M_\star$ obtained using the naive power counting, while in the right panels we show the limits on $M_\star$ obtained with the improved power counting.  The upper and lower panels present the limits obtained from $h \to $ inv and $Z \to$ inv, respectively.

We have included all the leading effect from the operators of eq. \eqref{eq:EFT} with a caveat for the ones appering in \eqref{eq:dim4}. Indeed, they do not show up in the naive power counting as they are 'renormalizable' and no power of $M_\star$ appears in the naive scaling.

We also highlight the regions in which the validity of the EFT computation is questionable: the horizontal regions show when $M_\star < 5 M_{h,Z}$ (light grey) or $M_\star < 10 M_{h,Z}$ (dark grey), depending on the case considered, while the oblique  regions refer to $R < 5$ (light blue) and $R < 10$ (lighter blue).

As it is clear from the left panels, when the naive power counting is used for the operators that do not contain the DM field strength and are, thus, not invariant under the dark $\UD$, the bounds become stronger as we move to lower values of $M$. Namely, the bound mass $M_\star$ becomes larger and larger. This is due to the $1/M^4$ enhancement discussed in sec.\,\ref{sec:stueckelberg} and it is cured once the improved power counting is used (see right panels). A particular behavior is displayed by the class of non-gauge invariant CP-odd operators, associated to the Wilson coefficients $(c_{\tilde{H}c},c_B^\prime,\,c_{HB}^\prime)$ involved in the Z boson decay. This is due to the presence of the Levi-Civita tensor in their definition. As can be seen from table \ref{tab:Zdecay}, in these cases the decay width has a weaker dependence on the DM mass when the naive power counting is used, scaling as $M^{-2}$. This behavior is dramatically affected when the choice on the power counting is modified, namely for $M \ll M_Z$ the decay width drops proportionally to $M^2$ when we apply the improved power counting. As a consequence, the bound on $M_\star$ becomes irrelevant for small masses, since the decay width vanishes in the $M\to 0$ limit. 

\begin{figure}[t]
\begin{center}
\includegraphics[width=0.5\textwidth]{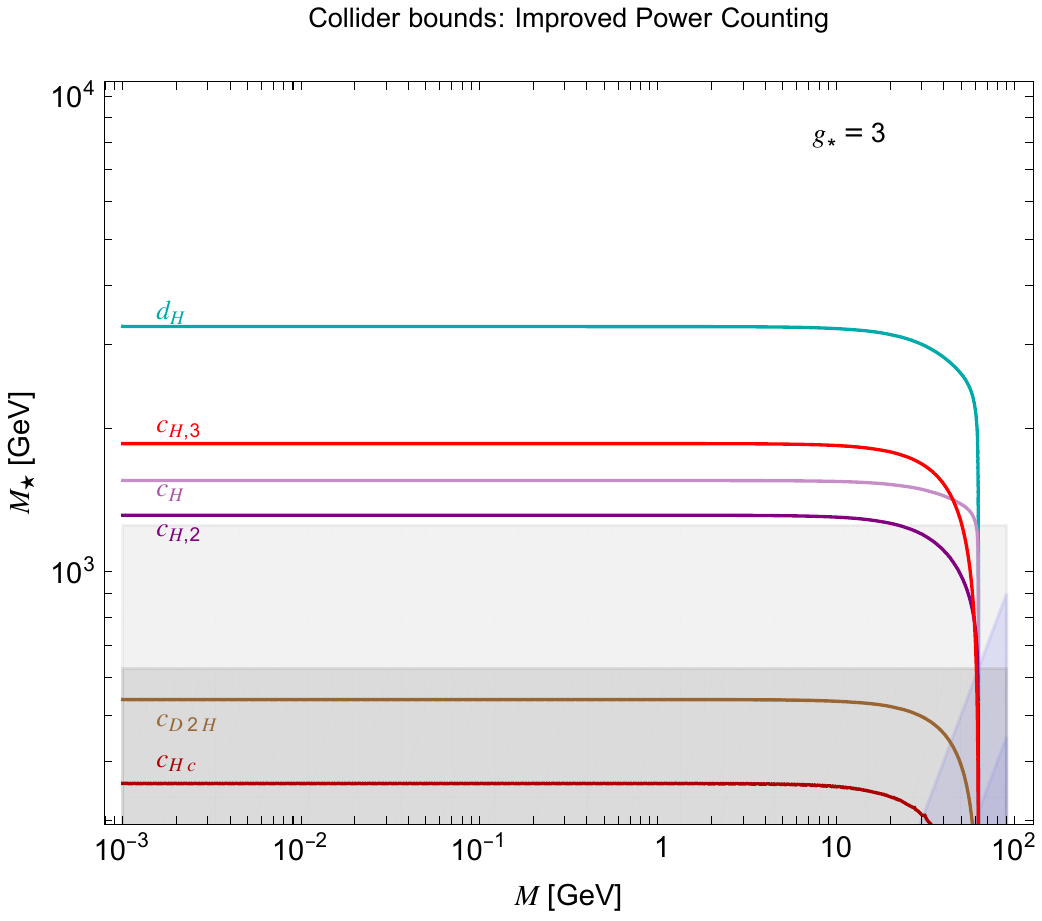}
\caption{Bounds on the parameters space $(M,\,M_\star)$ from invisible decay of $H$ and $Z$ bosons. Each line correspond to the Wilson coefficient of the EFT as in table \ref{tab:hdecay}. We show the constraints arising from one operator at a time, and the region below each curve is excluded. The results apply for $g_\star=3$ with all the dimensionless coefficients $d_i,\,c_i=1$ when the $improved$ power counting is used. Shaded regions correspond to regions of parameters space where the EFT validity breaks down: $M_\star\leq(5,10)M_h$ in grey and  $R\leq [5,10]$ in light blue.}
\label{fig:collider_summary}
\end{center}
\end{figure}

In general we observe that the bounds obtained using the improved power counting are less constraining than those obtained using the naive power counting due to the additional $M^2/M_\star^2\ll 1$ suppression. When using the improved power counting scheme, the choice $g_\star =1$ implies that the limits fall in a region in which the validity of the EFT is questionable (see figure \ref{fig:collider_summary}). As such, colliders limits are then mostly applicabile in EFT terms when the underlying model has some strong coupling.

The limits for different values of $g_\star$  can be easily obtained from the ones shown in Fig.\,\ref{fig:BRL} by an appropriate rescaling. For instance, consider the bound on $M_\star$ obtained switching on the operator $V_\mu \bar{V}^\mu |H|^4$, with coefficient $[1,\,M^2/M_\star^2] g_\star^4 c_H/M_\star^2$ as in eq.\,\eqref{eq:dim6}. The bound on $M_\star$ can be written as  \be\label{eq:Mstar_bound}
M_\star[M] \geq \left(M_\star\right[M])^{{\rm bound}, g_\star=1} \, \left\{ \begin{array}{lcl} g_\star^2 & ~~~ & {\rm (naive)}\\ g_\star & ~~~ & {\rm (improved),} \end{array} \right.
\ee
where $(M_\star[M])^{{\rm bound}, g_\star=1}$ is the experimental bound as a function of the DM mass obtained in a given power counting, that can be read from Fig.\,\ref{fig:BRL}. The different dependence on $g_\star$ in eq.\,\eqref{eq:Mstar_bound} is due to the different dependence of the decay widths on $M_\star$ in the two power counting schemes. For the remaining operators, it is now straightforward to repeat this reasoning considering the correct powers of $g_\star$ and to obtain the bound for any value of the coupling. 

This exercise shows that the bounds from invisible decays of $H$ and $Z$ are not so constraining as one might think. We show that this is especially true for a weakly coupled realization. From the right panels of figure \ref{fig:BRL} we see for example that also the 'Higgs portal' $c_H$ is not so constraining once the improved power counting is applied. This is also what happens in explicit realizations and we believe it is the right way to inspect EFT of dark photon dark matter.

\bigskip

\begin{figure}[t]
\begin{center}
\includegraphics[width=0.5\textwidth]{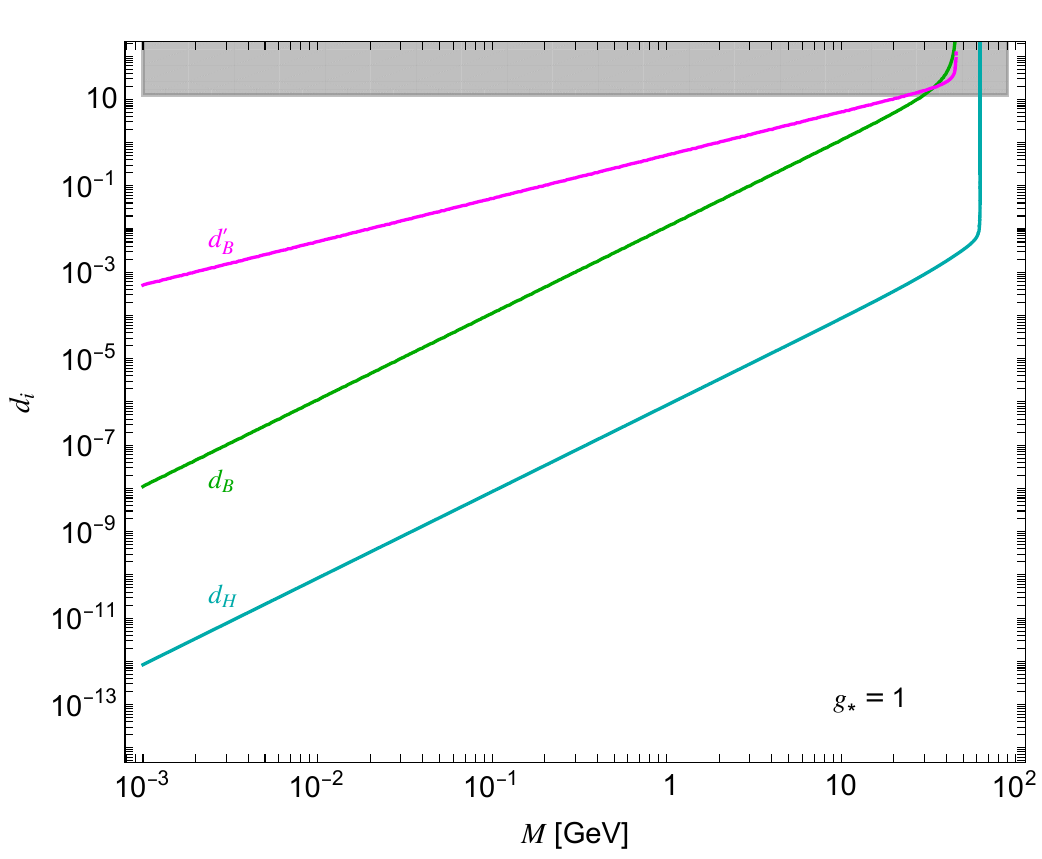}
\caption{Upper limits on the $(d_H,\,d_B,\,d_B^\prime)$ coefficients, corresponding to the Wilson coefficients of renormalizable operators in eq.\,\eqref{eq:dim4} when the naive power counting is used. The curves are obtained for $g_\star=1$ and represent the constraints obtained from Higgs and Z bosons invisible decay modes. 
}
\label{fig:BRC}
\end{center}
\end{figure}

As a final remark we show the bounds on the quartic couplings $d_H$ and $d_B^{(\prime)}$ when the naive power counting is used in Fig.\,\ref{fig:BRC}, where the constraints live in the plane ($M$, couplings). For $d_H$, this is the usual bound on the Higgs portal coupling \cite{Yu:2011by,Kumar:2015wya,Belyaev:2016pxe,Belyaev:2018pqr,Arcadi:2020jqf,Arcadi:2021mag,Aebischer:2022wnl}, while the $d_B^{(\prime)}$ coupling has been studied in the context of the coupling between vector DM and electromagnetic and weak multipoles\,\cite{Hisano:2020qkq,Chu:2023zbo}. In this case, the grey shaded region is drawn in correspondence of the benchmark value $d_i\geq 4\pi$, which is the typical value for breakdown of perturbativity and unitarity.

\paragraph{Summary of collider searches}~\\
Our procedure has identified a selection of operators that are mostly constrained by invisible decays of $H$ and $Z$.  As discussed, the bounds are not important  (fig. \ref{fig:BRL}) for a theory of weakly coupled complex dark photon $(g_\star=1)$ when we apply the correct power counting solution i.e. the \textit{improved} one (see sec. \ref{sec:power_counting}) since they apply in a region beyond the EFT validity. 

In figure \ref{fig:collider_summary}, in which we show the bounds for a moderately strong bound $g_\star = 3$, we can appreciate that only the operators connected with the Higgs invisible decay mode happen to escape the regions where the validity of the EFT breaks down as the requirement on the separation of scales fails. In this case, the well known Higgs portal operator $d_H$ (\textit{dark cyan} line) sets the largest lower bound on the new physics scale $M_\star$ and would therefore be relevant for the effective theory phenomenology. Other relevant exclusion regions in this scenario are given by $(c_{H},\,c_{H,2},\, c_{H,3})$ which are the dimension six Higgs portal, the CP-even and CP-odd field strength operators, respectively. As a reference, we also show that the strongest bound from $Z\to\text{inv}$ associated to the \textit{dark red} (labelled $c_{\tilde{H}c}$) still cannot overcome the EFT validity region.

\subsection{Direct Detection}\label{subsec:DD}

\begin{figure}[p]
\centering
\includegraphics[width=0.4\textwidth]{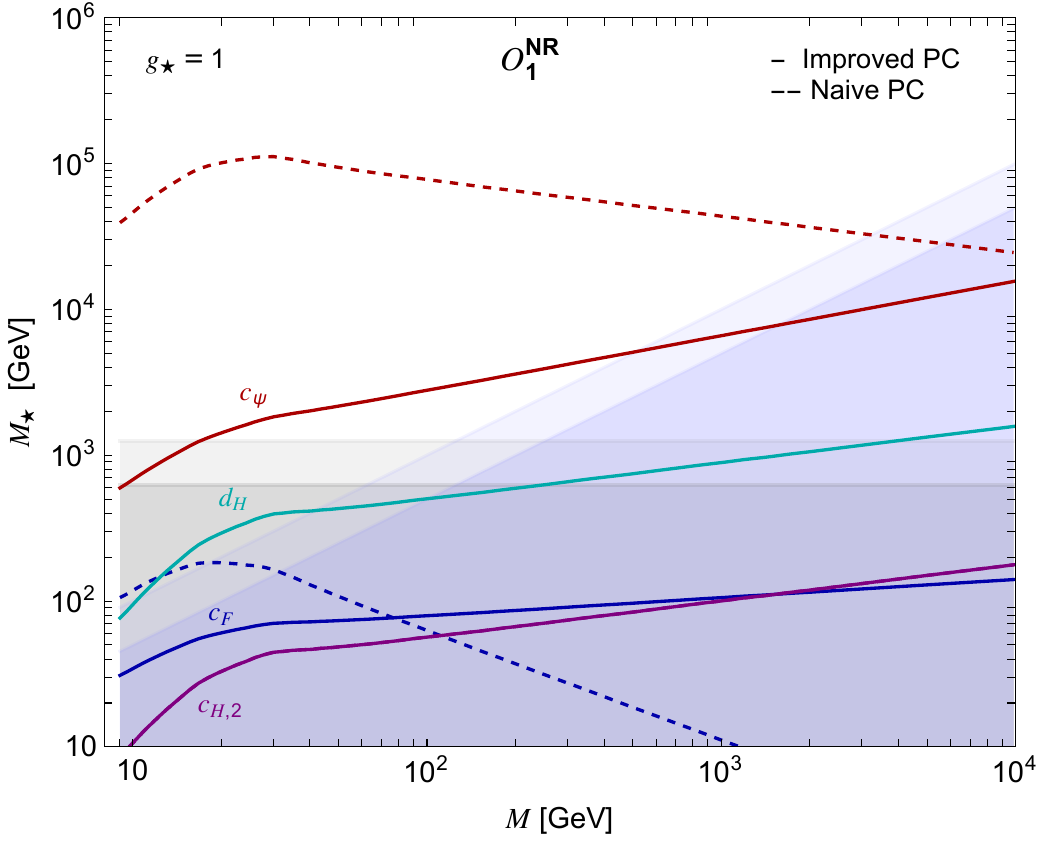}
\includegraphics[width=0.4\textwidth]{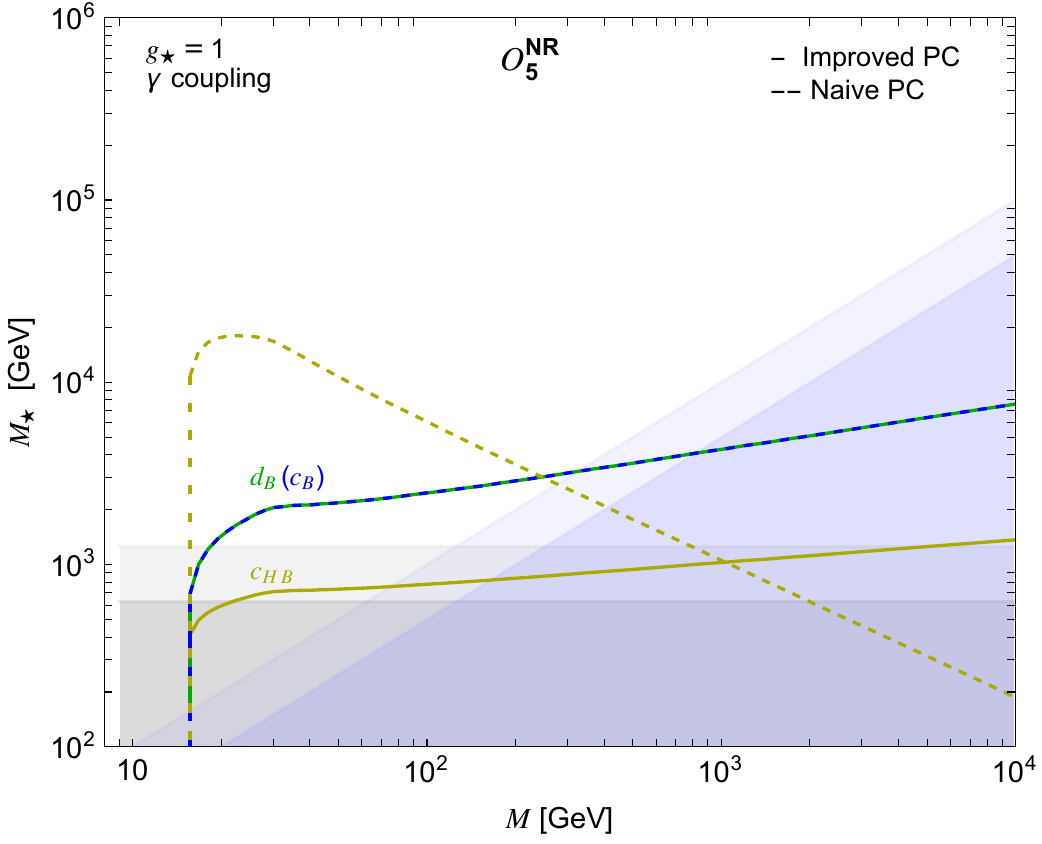}\\
\includegraphics[width=0.4\textwidth]{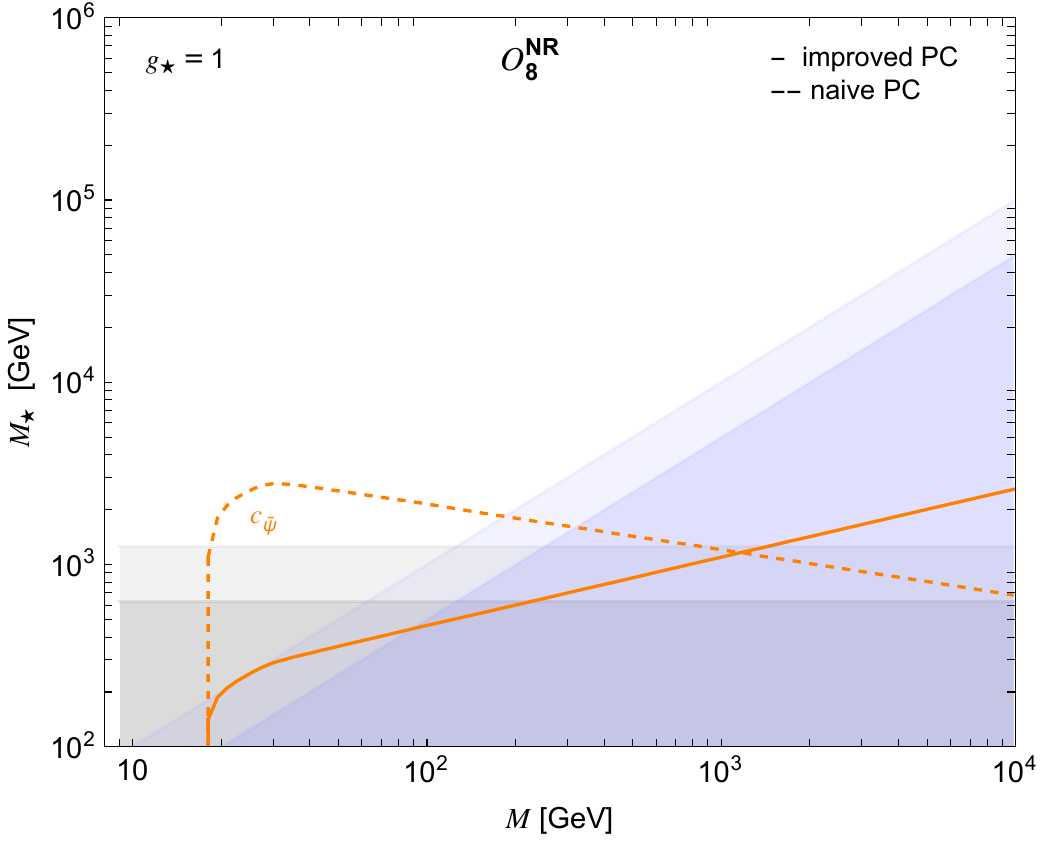}
\includegraphics[width=0.4\textwidth]{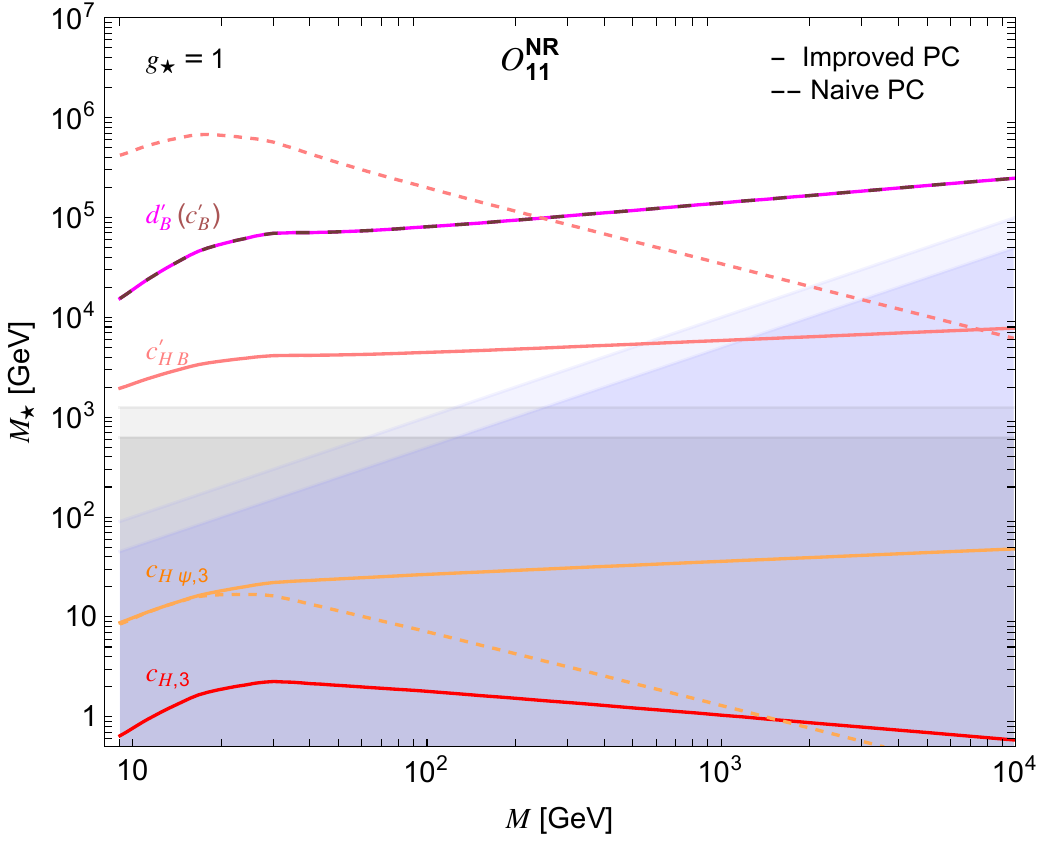}\\
\includegraphics[width=0.4\textwidth]{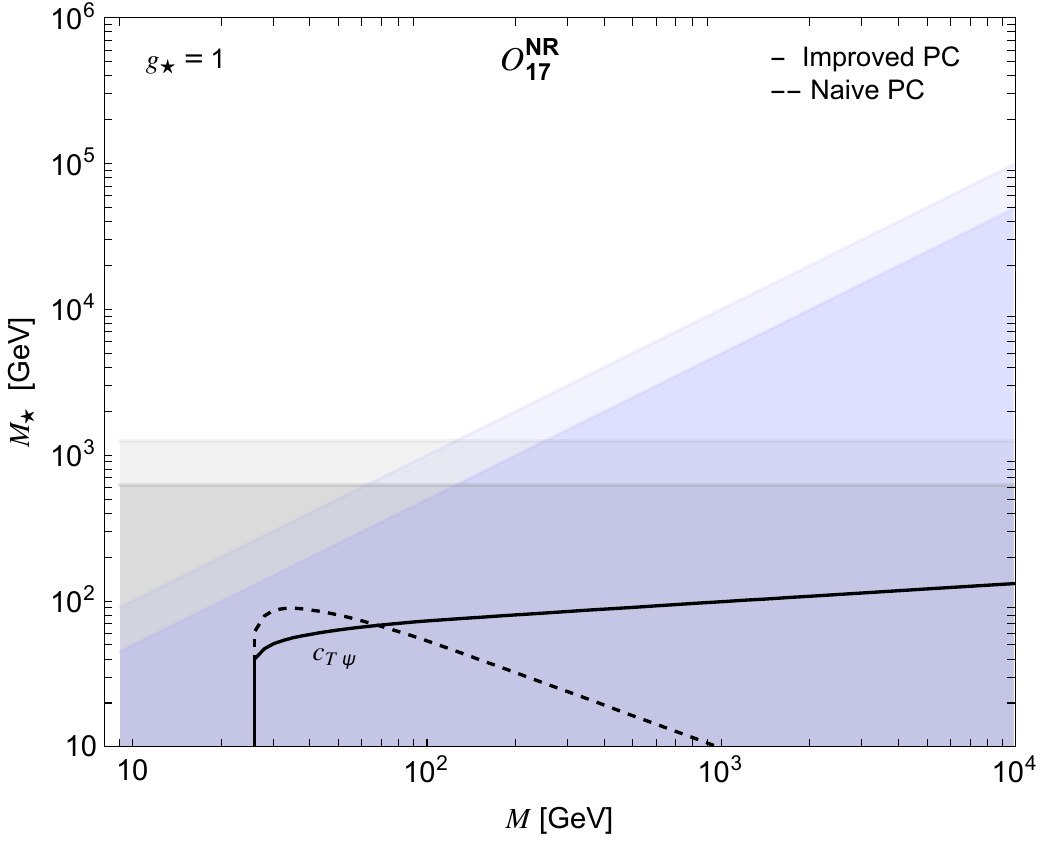}
\includegraphics[width=0.4\textwidth]{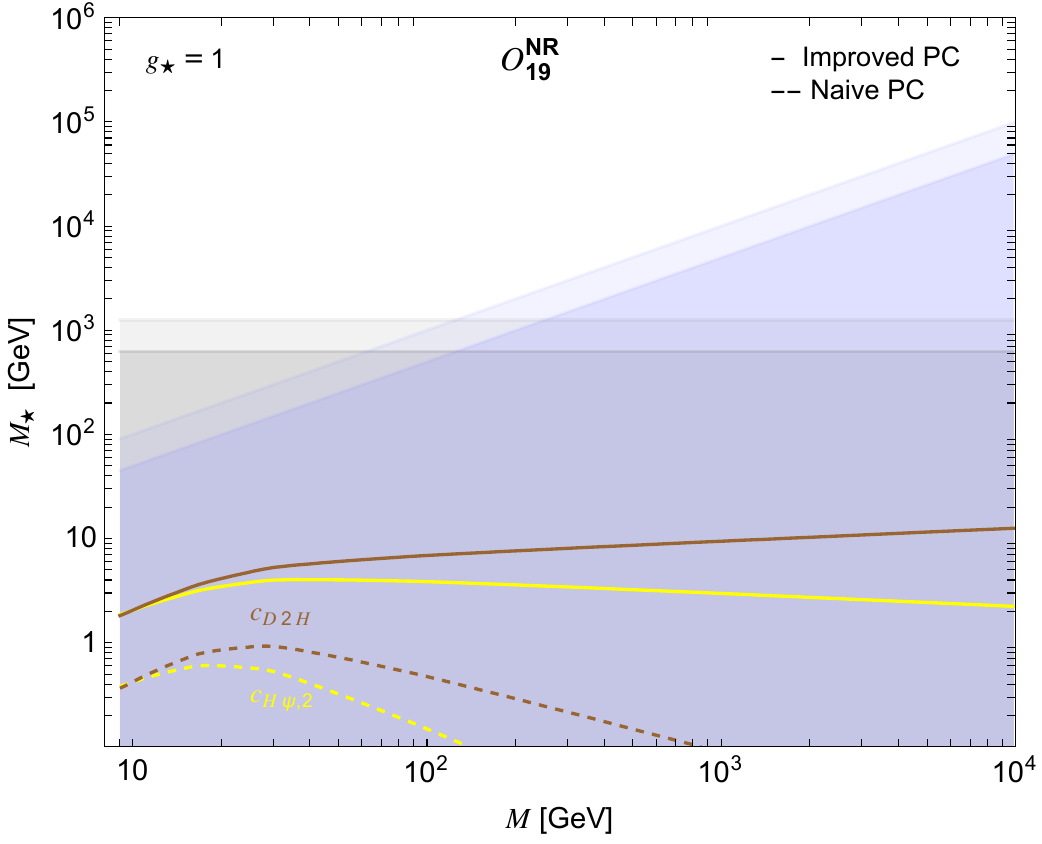}
\caption{Lower bounds on the new physics scale $M_\star$ derived for each operator $\Op_i^\textup{\tiny{NR}}$ contribution to our "reduced" SI NREFT describing a complex dark photon undergoing elastic scattering off nuclei. We compare the constraints derived adopting both the naive (\textit{dashed lines}) and the improved (\textit{solid lines}) power counting solutions. All curves are plotted for the benchmark value $g_\star=1$. \textit{Grey} and \textit{lightblue} shaded regions correspond to parameters space patches where the EFT validity is questionable, i.e. when $M_\star\leq(5,10)M_h$ and $R=M_\star/M\leq(5,10)$ respectively. }
\label{fig:DD bounds on Mstar}
\end{figure}

In this section we describe how direct detection experiments can put bounds on the parameter space of the EFT we are considering. Despite direct detection experiment work at low energy transfer, therefore in a regime where the non-renormalizable nature of massive vectors does not appear, we find that the power counting developed in the previous section still provides a suppression to otherwise unphysical large effects in the rates. The bounds derived here connects with the discussion of sec. \ref{sec:matching}.

As already mentioned in that section, we consider only spin-independent processes, which should impose the stronger constraints on our model. The relevant Lagrangian we consider is the one in eq. \eqref{eq:NREFT}, with Wilson coefficients given in eq. \eqref{eq:matchingNR}. As  can be seen with our derivation, the EFT Lagrangian of eq.\,\eqref{eq:EFT} generates only six non-relativistic spin-independent operators
\be
\mathrm{direct\ detection}:\,\quad (\Op_1^\textup{\tiny{NR}},\Op_5^\textup{\tiny{NR}},\Op_8^\textup{\tiny{NR}}, \Op_{11}^\textup{\tiny{NR}},\Op_{17}^\textup{\tiny{NR}},\Op_{19}^\textup{\tiny{NR}}),
\ee
and we refer again to table \ref{tab:NRop} for the definitions. We focus on these in the remainder of this section. Naively one would expect that $\Op_1^\textup{\tiny{NR}}$ would give the strongest constraint, but the results will be radically different when low-momentum enhancement are considered.

In order to set the limits coming from direct detection experiments, we use the following approximate procedure. We use the 95\% C.L. exclusion obtained by the LZ experiment on the DM-nucleon spin-independent cross section\,\cite{LZ:2022lsv} and plug it inside eq.\,\eqref{eq:Nth} in appendix \ref{app:xsectionDD} to obtain the number of experimentally excluded events $N_{\rm exp}(M)$ as a function of the DM mass. For this computation, we use the Helm form factor as nuclear response function representative of the spin-independent search. For each of the Wilson coefficients appearing in eq.\,\eqref{eq:dim4} and eq.\,\eqref{eq:dim6}, we then compute the number of events $N_{model}(M; \left\{\wc_i, g_\star, M_\star\right\})$, following all the matching procedure and the computation of the cross section as explained in appendix \ref{app:xsectionDD}. We obtain the excluded region imposing
\be
N_{\rm model}(M; \left\{ \wc_i, g_\star, M_\star\right\}) < N_{\rm exp}(M)\,.
\ee
More details on the computation of the number of events at direct detection experiments can be found in appendix \ref{app:xsectionDD}.

\bigskip

Our results are presented in Fig.\,\ref{fig:DD bounds on Mstar}, where we show a comparison between the limits obtained using the naive (dashed lines) and improved (solid lines) power counting in the plane $(M,M_\star)$. Each panel shows one of six different non-relativistic operators generated when matching onto the NREFT of eq.\,\eqref{eq:NREFT} as discussed in sec.\,\ref{sec:matching}. Shaded regions are the same as in Fig.\,\ref{fig:BRL} and show the regions in which the EFT is not a valid expansion anymore. In all the plots the dimensionless coefficients $d_i,\,c_i$ are set to unity, as well as the dark coupling $g_\star$. For each line $M_\star(M)_{\rm excl.}$, the excluded region stays below that line.

In discussing our surveys of bounds, two main points have to be considered. 
First, we see that the improved power counting strongly relaxes the bound in the EFT regime (when $R\gg1$), since in all the plots the solid line stay below the dashed line of corresponding color.
Second, some non-relativistic operators are enhanced by inverse powers of $q^2$, which may overcome the suppression of $q$ and $v$ at the numerators. This fact allows us to explain why the bound on $M_\star(M)$ from $\mathcal{O}_1^\textup{\tiny{NR}}$ is comparable to the bound arising from $\mathcal{O}_{5,11}^\textup{\tiny{NR}}$.
The operator $\mathcal{O}_1^\textup{\tiny{NR}}$ is the only NR interaction not suppressed by powers of DM velocity or exchanged momentum. However, we obtain physically significant bounds also from $\Op_{11}^\textup{\tiny{NR}} = i (\vec q\cdot\SV)/m_N$ (that carries one factor of the exchanged momentum $\vec q$), and from $\Op_5^\textup{\tiny{NR}} = i \SV \cdot (\vec{q} \times \vec{v}_\perp)/m_N$ (further suppressed by one factor of $\vec v_\perp$). These results could appear quite surprising, but they are easily explained remembering that, despite the $\sim q,\,\sim q v_\perp$ suppression, both Wilson coefficients of these NR operators receive leading order contributions in a $q^n$ expansion from the magnetic and electric "dipole-like" interactions 
$$\{V_{[\mu}\bar V_{\nu]}B^{\mu\nu}\,,\,V^{}_{[\mu\rho}\bar V^\rho_{\,\,\,\nu]}B^{\mu\nu},V_{[\mu}\bar V_{\nu]}\tilde B^{\mu\nu}\,,\,V^{}_{[\mu\rho}\bar V^\rho_{\,\,\,\nu]}\tilde B^{\mu\nu}\},$$
that are labelled by coefficients $d_B^{(\prime)},\,c_B^{(\prime)}$ as in eq. \eqref{eq:d_coeff} and table \ref{tab:dim6_h_counting}.
This results in a $1/\vec q\,^2$ enhancement factor typical of the long range interaction at low exchanged momentum, so that we obtain a significant bound in a region in which the EFT is valid. In particular we point out that complex dark photon DM is the only scenario where a dimension-4 operator - although cured by the improved power counting - can give rise to an electric dipole moment interaction, corresponding to $q^{-2}\Op_{11}^\textup{\tiny{NR}} = q^{-2}i (\vec q\cdot\SV)/m_N$, therefore giving a low-momentum enhancement which is unsuppressed by the DM velocity (contrary to the magnetic dipole interaction, which is still suppressed by $\vec{v}_\perp$).

\begin{figure}[tb]
\centering
\includegraphics[width=0.5\textwidth]{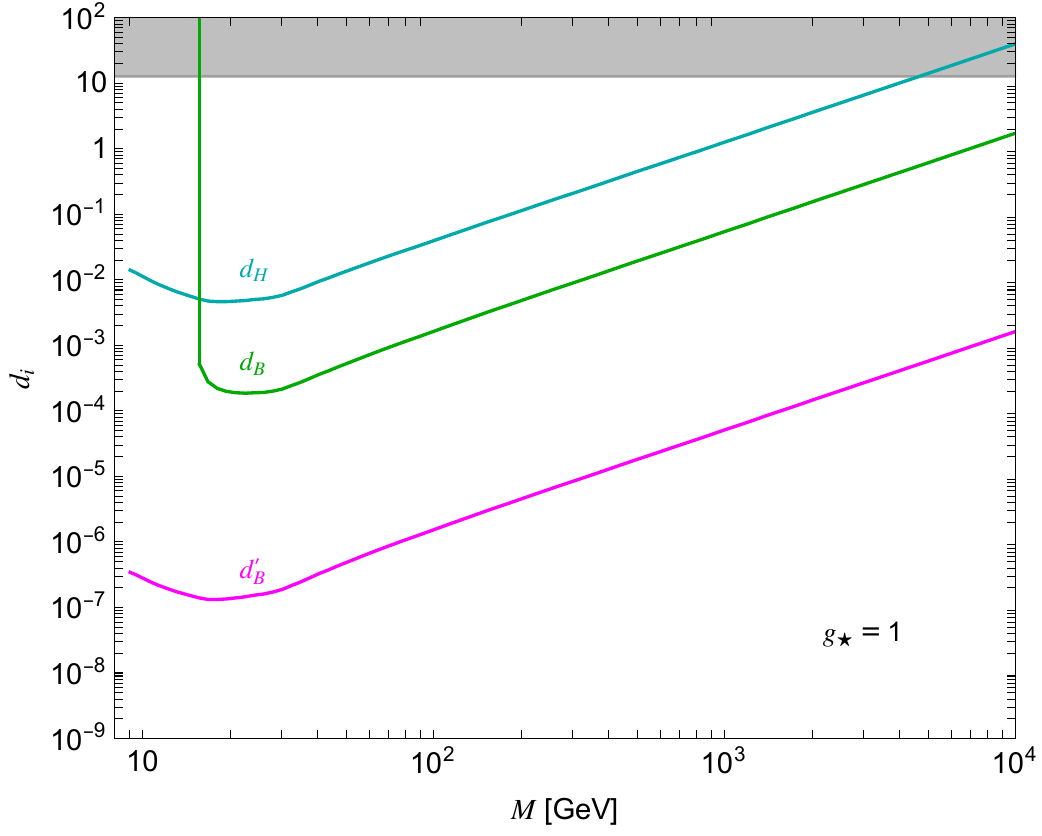}
\caption{Upper limits on the $(d_h,\,d_B,\,d_B^\prime)$ coefficients, corresponding to the Wilson coefficients of renormalizable operators in eq.\,\eqref{eq:dim4} when the naive power counting is used. The curves are obtained for $g_\star=1$ and represent the constraints obtained from spin-independent direct detection at LZ. }
\label{fig:lambdaiDD}
\end{figure}

In terms of the naive power counting rule, we can impose upper limits on the coefficients of the four operators of \eqref{eq:dim4}. The results are shown in Fig. \ref{fig:lambdaiDD}.

Before concluding, it is worth emphasizing that half of the bounds on the NR operators do not extend over the whole mass range investigated by the LZ experiment, with the curves terminating abruptly, rather than smoothly. This feature can be traced to their $\vec q,\,\vec v_\perp$ dependence. These enter the calculation of the total number of theoretically expected events through the velocity (astrophysical uncertainy) and recoil energy (experimental input) integrated form factor discussed in eq. \eqref{eq:integrated F} of Appendix \ref{app:xsectionDD}, which is (partially) responsible of the shape and responsible of the fact that they close earlier than the experimental window.

\begin{figure}[t]
\begin{center}
\includegraphics[width=0.55\textwidth]{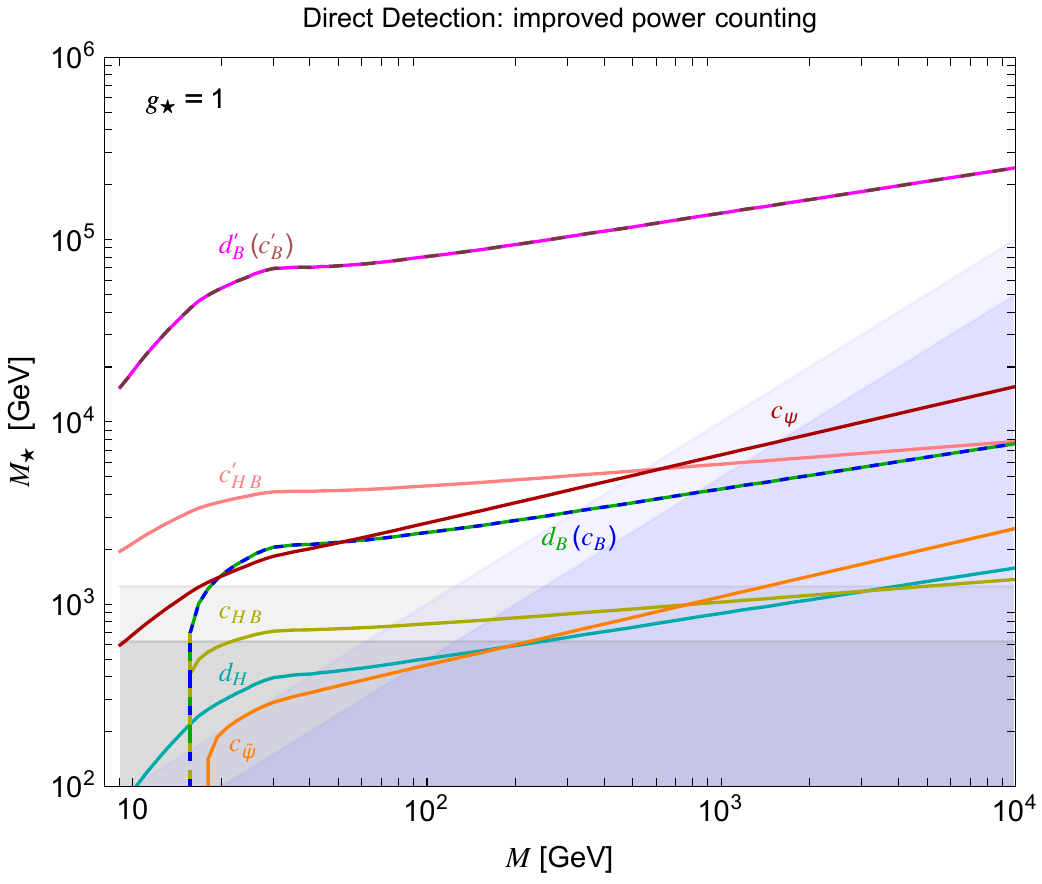}
\caption{
Bounds on the parameters space $(M,\,M_\star)$ from spin independent direct detection at LZ. Each line correspond to the Wilson coefficient of the EFT as in table \ref{tab:NRop}. We show the constraints arising from one operator at a time, and the region below each curve is excluded. The results apply for $g_\star=1$ with all the dimensionless coefficients $d_i,\,c_i=1$ when the $improved$ power counting is used. Shaded regions correspond to regions of parameters space where the EFT validity breaks down: $M_\star\leq(5,10)M_h$ in grey and  $R\leq [5,10]$ in light blue.}
\label{fig:DD_summary}
\end{center}
\end{figure}

\paragraph{Summary of direct detection}~\\
Our procedure has identified a selection of operators that are mostly constrained by direct detection. In figure \ref{fig:DD_summary} we focus on the resulting exclusion curves $M_\star(M)$ obtained for a weakly coupled dark sector $g_\star=1$ when the correct behaviour of effective operators under gauge transformations is restored applying the \textit{improved} power counting (see sec. \ref{sec:power_counting}). All curves correspond to $d_i,\,c_i=1$.

We notice that the exclusion of a few interactions (as for example the Higgs portal $d_H$, \textit{dark cyan} line) fall in a region beyond the EFT validity. We conclude that Higgs portal phenomenology is not so relevant for vector DM (see however \cite{Arcadi:2021mag}).

However, other operators receive meaningful exclusion limits in a region where the EFT is fully valid. This is the case for dipole-like operators corresponding to $d_B^{(\prime)},c_B^{(\prime)}$ and $c_{HB}^\prime$. They provide the strongest lower bounds on the new physics scale $M_\star(M)$ even for moderate couplings $g_\star=1$. This was already noticed in \cite{Hisano:2020qkq}, that used previous Xenon data. For complex vectors, the most interesting channel is the one mediated by electric-dipole interaction with the photons $\Op_{\mu\nu}^A \tilde F^{\mu\nu}$ corresponding $d'_B$. In the NREFT it receives a $1/\vec q^{\,2}$ enhancement at low momentum transfer -- typical of long-range interactions mediated by the photon -- and no $v$ suppression that set $M_\star(M)\gtrsim10^4-10^5\,\GeV$ throughout the investigated mass window. It gives a strong bound even with the improved power counting.

\subsection{Massless Stueckelberg case}
\begin{figure}[tb]
\centering
\includegraphics[width=0.55\textwidth]{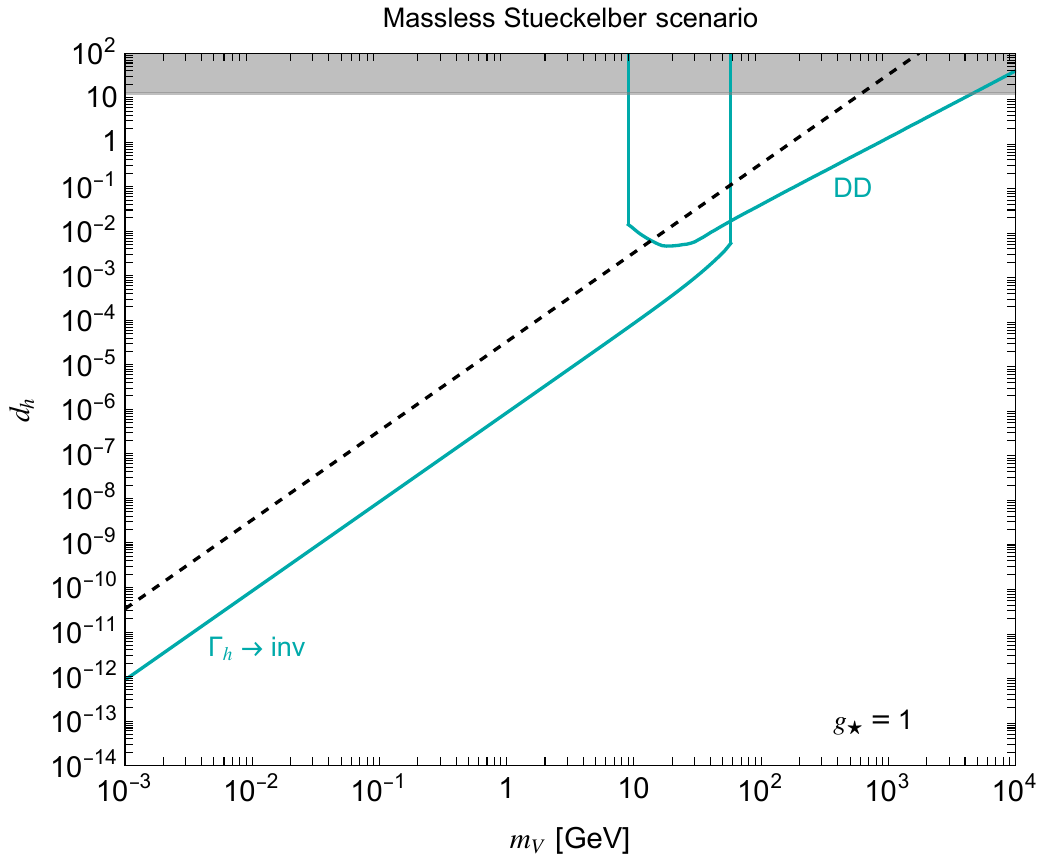}
\caption{Limits coming from the Higgs invisible BR and DD searches on the massless Stueckelberg scenario. The dashed line represents the values of parameters in which eq. \eqref{eq:DP_vev_mass} is satisfied, namely in which the dark photon mass is completely generated by contact interactions between DM and SM fields (more specifically, by the $d_H$ coefficient).}
\label{fig:baremass}
\end{figure}
In section\,\ref{sec:DP_physical_mass} we briefly discussed the possibility that the physical complex dark photon mass is completely generated by the EFT operators, once we set $M=0$. Considering only the leading effect coming from the Wilson coefficient $\lambda_H$, we would have that the dark photon mass equals
\be\label{eq:DP_vev_mass}
 m_V^2=d_H\, g_\star^2 \frac{\vev^2}{2},
\ee
i.e. it is fixed by the coefficient $d_H$. In this case, we can then study when collider and direct detection bounds can place constraints on such a scenario. We show our results in fig.\,\ref{fig:baremass}, where the dashed line represents the dark photon mass of eq. \eqref{eq:DP_vev_mass}. The experimental bounds obtained from the Higgs invisible BR and from direct searches at LZ are shown together. As we can see, the limit coming from the Higgs decay width exclude this scenarios for masses below $M_h/2 \simeq 63$ GeV. On the other hand, the limits coming from the LZ experiment exclude the scenario for masses above $(14-15)$ GeV, meaning that this simple scenario is completely excluded. We observe that the conclusion is independent on the value of $g_\star$, since the experimental limits and the dark photon mass scale in the same way once we allow for $g_\star \neq 1$.

\section{Sketching possible UV completions}\label{sec:UV}
In this section we present two explicit UV complete models that generate, under some assumptions about hierarchy of vacuum expectation values, some of the operators that appear in the low energy EFT of a complex dark photon. In our discussion, we pay special attention to the requirements necessary to generate the improved and the naive power counting described in Sec.\,\ref{sec:power_counting}. 

\subsection{Model \bf{\rm{I}}:  SU(2)$\times$ U(1)}\label{sec:model1}
The first model we present is based on a SU(2)$\times$ U(1) dark gauge symmetry (see also Refs.\,\cite{Hisano:2020qkq,Chiang:2013kqa,Davoudiasl:2013jma,Choi:2017zww,Choi:2019zeb,Catena:2023use}), completely broken by the vacuum expectation values (vevs) of two scalars transforming as $\phi \sim (\bm{d},q)$ and $\varphi \sim (\bm{1}, q')$. Here $\bm{d}$ denotes a generic dimension-$d$ representation of SU(2) and $q$, $q'$ are the U(1) charges. Both scalars are complete singlets under the SM gauge group. We denote by $V^a$ and $X$ the gauge bosons of SU(2) and U(1), respectively, with corresponding gauge couplings $\gd$ and $\gdp$. The DM candidate $V$ and its antiparticle are given by the combinations $V = (V^1 - i V^2)/\sqrt{2}$ and $\bar V = (V^1+i V^2)/\sqrt{2}$.  As we are now going to show, the EFT generated integrating out the heavy scalar and vector states will include the Higgs portal interactions, as well as vector $\times$ vector and tensor $\times$ tensor operators. 

For what concerns the generation of effective operators via scalar exchange, the relevant interactions are contained in the following potential
\be\label{eq:model1}
V(\phi,\varphi,H) =-\mu^2_\phi|\phi|^2+\lambda_\phi|\phi|^4-\mu^2_\varphi|\varphi|^2-\lambda_\varphi|\varphi|^4-\lambda_1|\phi|^2|\varphi|^2-\lambda_2|\phi|^2|H|^2-\lambda_3|\varphi|^2|H|^2\,\,,
\ee 
that contains all the terms involving $\phi$, $\varphi$ and their interactions with the Higgs doublet $H$. In what follows, we will always consider scenarios for which 
\be
\langle \varphi \rangle \gg \langle \phi \rangle, \langle H \rangle ,
\ee
in such a way that the mixing between $\phi$ and $\varphi$, that generates the \textit{Higgs portal}, is given by
\be\label{eq:scalarmix}
\theta_{\varphi\phi}\simeq\frac{\lambda_1}{\lambda_\varphi}\frac{\braket{\phi}}{\braket{\varphi}}\,\,.
\ee

Similarly, the kinetic mixing $\varepsilon$ between $U(1)_\text{Y}$ and U(1) induces a mixing between the dark gauge bosons $V^3,\,X$ and the SM $Z_\textup{\tiny{SM}}$ and generates a \textit{vector portal} between the dark and visible sector. After symmetry breaking, we can write the dynamical fields $V^3$, $X$ and $Z_\textup{\tiny{SM}}$ in terms of the mass eigenstates $Z^\prime$, $D$ and $Z$ as follows:
\begin{align}\label{eq:vectormix}
&\begin{cases}
V^3\simeq Z^\prime+\theta_{XV^3}\,D+\dots\\
X\simeq D-\theta_{XV^3}\,Z^\prime-\theta_{XZ}\, Z_{\tiny{SM}}+\dots\\
Z\simeq Z_\textup{\tiny{SM}}+\theta_{\tiny{XZ}}\,D+\dots
\end{cases}
\begin{cases}
\theta_{XV^3}\simeq\frac{q\,g^\prime_\textup{\tiny D}}{m\sqrt{1-\varepsilon^2}g_\textup{\tiny D}}\frac{M^2}{M_X^2}, \\
\theta_{XZ}\simeq\frac{\varepsilon g^\prime}{g_Z\sqrt{1-\varepsilon^2}}\frac{M^2_Z}{M_X^2}\,,\\
\end{cases}
\end{align} 
where we have written the expansion for the mixing angles. The vector masses are
\begin{align}\label{eq:vectormasses}
&M=\gd \langle \phi \rangle \sqrt{\ell(\ell+1) - m^2}, &M_{Z^\prime}&\simeq m\gd\braket{\phi}, \\
&M_D\simeq\frac{\gdp q^\prime}{\sqrt{1-\varepsilon^2}}\braket{\varphi}, &M_Z &\simeq M_{Z_\textup{\tiny{SM}}},
\end{align}
where $\ell = (d-1)/2$ and $m = -\ell, \dots, +\ell$ are the usual $SU(2)$ quantum numbers ($m$ labels the direction along which the vev is aligned \cite{Nomura:2020zlm}).

We now focus on the generation of the EFT. For the moment, we do not assume any hierarchy between $\braket{\phi}$ and $\braket{H}$, so that the only heavy states are those whose masses are $\mathcal{O}(\braket{\varphi})$, \textit{i.e.} $D\simeq X$ and $\varphi$. The operator $V_\mu \bar V^\mu |H|^2$ is generated by the scalar portal once $\varphi$ is integrated out. We obtain
\be\label{eq:varphi_exchange}
\M_{V\bar V H^\dag H}^{(\varphi)} = g_{\varphi V \bar V} \frac{1}{M_\varphi^2} g_{\varphi H^\dag H}\epsilon_V \epsilon_{\bar V} \sim \frac{\lambda_1 \lambda_3}{\lambda_\varphi} \frac{M^2}{M_\varphi^2}\epsilon_V \epsilon_{\bar V}\,\,, \quad\quad \text{with}\, \frac{\lambda_1 \lambda_3}{\lambda_\varphi} \sim g_\star^2,
\ee
where $\epsilon_V\epsilon_{\bar V}$ denote, schematically, the DM polarization vectors and we have used that the coupling between $\varphi$ and the $V\bar V$ pair goes like $g_{\varphi V\bar V} \sim (M^2/\braket{\phi}) \theta_{\varphi \phi}$, while the coupling between $\varphi$ and $H^\dag H$ scales as $g_{\varphi H^\dag H} \sim \lambda_3\braket{\varphi}$. As we see, the Wilson coefficient obeys the ``improved'' power counting of Sec. \ref{sec:stueckelberg}, once we identify $M_\star \sim M_\varphi$.\\
Turning now to the current $\times$ current operator $J_V^\mu J^B_\mu$, this is generated via the vector portal once we integrate out the heavy $X$ state. We obtain (up to numerical factors)
\be\label{eq:Xexchange}
\M_{V \bar V J_B}^{(X)} = g_{XV\bar V} \frac{1}{M_X^2} g_{X J_B}\frac{1}{\sqrt{1-\varepsilon^2}}  \epsilon_V \epsilon_{\bar V} \sim \frac{g_\textup{\tiny D}^\prime\, g'\, q\,\epsilon}{\gd} \frac{M^2}{M_X^4} \epsilon_V\epsilon_{\bar V}\,\,, \quad\quad \text{with} \, \frac{g_\textup{\tiny D}^\prime\, g'\, q\,\epsilon}{\gd} \sim g_\star^2, 
\ee
where have used $g_{XV\bar V} \sim \gd \theta_{\scriptstyle{V^3X}}$, $g_{\scriptstyle{XJ_B}} \sim g' \epsilon$, while the mixing angle can be found in eq.\,\eqref{eq:vectormix}. Once again, we appreciate that the naive power counting $1/M_\star^2 = 1/M_X^2$ is corrected by the factor $M^2/M_\star^2$, \textit{i.e.} the ``improved'' power counting applies. Qualitatively similar results hold when we make the further assumption and consider $\braket{\varphi}\gg \braket{\phi} \gg \braket{ H}$. For instance, if we consider the amplitude with a (now heavy) $\phi$ exchange, we obtain
\be\label{eq:phi_exchange}
\M_{V\bar V H^\dag H}^{(\phi)} = g_{\phi V \bar V} \frac{1}{M_\phi^2} g_{\phi H^\dag H} \epsilon_V \epsilon_{\bar V}\sim g_D^2 \frac{\lambda_2}{\lambda_\phi} \sqrt{\ell(\ell+1) - m^2} \epsilon_V \epsilon_{\bar V} \sim \frac{M^2}{M_\phi^2} \lambda_2 \epsilon_V \epsilon_{\bar V}\quad \quad \text{with}\, \lambda_2\, \sim g_\star^2 .
\ee
Once more, we obtain the improved power counting we have advocated for in the text. 

We conclude observing that the SU(2) non-abelian kinetic term contains, in addition to the $V^3_\mu J_V^\mu$ coupling between $V^3$ and the DM vector current, also the tensor coupling $V^3_{\mu\nu} V^{[\mu}\bar V^{\nu]}$. This means that the argument outlined above for the current $\times$ current operators can be immediately extended to the tensor $\times$ tensor operators generated through this term -- \eg the dipole-like contact term in the effective theory of eq.\,\eqref{eq:dim4} --  with the same conclusions about the power counting applying.

\subsection{Model \bf{\rm{II}}: $SU(2)_L \times SU(2)_R \times \mathbb{Z}_2$} \label{sec:model2}
The second model we consider is based on a $SU(2)_L \times SU(2)_R \times \mathbb{Z}_2$ gauge symmetry. This time, the DM candidate only interacts with the visible sector via a Higgs portal since, being the dark gauge symmetry non-abelian, there is no kinetic mixing between the dark and visible sectors forbidding any interaction mediated by vector bosons. For concreteness, we denote by $V_{L,R}^a$ and $g_\textup{\tiny{L,R}}$ the gauge bosons and couplings of $SU(2)_{L,R}$. The symmetry is completely broken in two steps: first to $U(1)_L \times U(1)_R\times \mathbb{Z}_2$ by the vev of two triplets, $\Sigma_L \sim (\mathbf{3}, \mathbf{1})$ and $\Sigma_R \sim (\mathbf{1}, \mathbf{3})$; then to $\mathbb{Z}_2$ by the vev of two doublets, $\phi_L \sim (\mathbf{2}, \mathbf{1})$ and $\phi_R \sim (\mathbf{1}, \mathbf{2})$. The $\mathbb{Z}_2$ symmetry acts as
\be
(\mathbb{Z}_2) ~~~~~~~~V_L^a \leftrightarrow V_R^a, ~~~ \Sigma_L \leftrightarrow \Sigma_R, ~~~ \phi_L \leftrightarrow \phi_R.
\ee
Invariance under dark parity transformation forces the equality between the gauge couplings $g_L = g_R \equiv g_d$, between the triplets vevs $\braket{\Sigma_L} = \braket{\Sigma_R} \equiv \braket{\Sigma}$ and between the doublets vevs $\braket{\phi_L} = \braket{\phi_R}\equiv \braket{\phi}$. The two-step symmetry breaking described above can be achieved assuming a hierarchy between dark scalar vevs $\braket{\Sigma}\gg \braket{\phi}$ and that the triplets vevs are aligned along the $\sigma^3/2$ generator. Under these hypothesis, the states $V_{L,R}^\pm = (V_{L,R}^1 \mp i V_{L,R}^2)/\sqrt{2}$ and the scalar triplets have $\mathcal{O}(\braket{\Sigma})$ masses, while $V_{L,R}^3$ and the scalar doublet are much lighter, with masses of order $\mathcal{O}(\braket{\phi})$. However, because of the $\mathbb{Z}_2$ symmetry, $V_L^3$ and $V_R^3$ are degenerate, so that we can identify our DM candidate via $V = (V_L^3 - i V_R^3)/\sqrt{2}$. Unlike what happened in the SU(2)$\times$ U(1) model, we have now successfully obtained a hierarchy between the mass of the DM and of the remaining states in the gauge multiplet without invoking large $SU(2)$ representations. The price we pay is the absence of the vector portal. The scalar portal operator  is now obtained integrating out the heavy triplets instead of $\varphi$. The amplitude we obtain is the same as in eq.\,\eqref{eq:varphi_exchange}, with the identification $M_\varphi \to M_\Sigma$ and new, appropriate quartic coupling. Also in this case we thus have a situation in which the ``improved'' power counting is valid.

\section{Conclusions}\label{sec:conclusions}
In this work, we have studied a scenario in which the dark matter candidate is a massive complex vector $V_\mu$, dubbed \textit{complex dark photon}. We assume that the spin-1 DM particle is the only low energy remnant of a more complex heavy dark sector with typical mass $M_\star$ that, once integrated out, generates a set of effective operators that can be studied phenomenologically. The stabilization of the DM candidate is achieved via an accidental dark $\UD$ symmetry which survives at low energy. 

An essential point that must be highlighted is the fact that, in the EFT of a massive vector field, the power counting of the operators is subtler than the one of theories with spin-0 or spin-1/2 DM candidates. This is because the free theory of a massive Stueckelberg field is renormalizable, but this renormalizability is lost in general once interactions are turned on. In practice, this means that the theory may cease to be valid at an energy parametrically smaller than the naive cutoff $M_\star$. To correct this behavior and reinstate $M_\star$ as cutoff of the EFT, we introduced the so-called ``improved power counting'', in which we replace $V_\mu \to (M/M_\star) V_\mu$. 

We then turned to the main point of the paper, i.e. the construction of the EFT Lagrangian at the scale $M_\star$, considering operators up to dimension 6 and paying particular attention to the elimination of redundant operators. With this information, we studied the effective theories obtained at lower energies, having in mind applications in direct detection experiments. We first integrated out SM particles at the electroweak scale (Higgs, $W$ and $Z$ bosons and the top quark). We then moved to energies of order 1 GeV and matched the operators to a single-nucleon relativistic EFT. Finally, we matched onto the non-relativistic theory that can be used to compute nuclear response functions and, ultimately, the number of events expected at direct detection experiments.

With all this information at our disposal, we finally turned to the phenomenological analysis, with the aim of putting limits on the Wilson coefficients of the operators defined at the scale $M_\star$. We considered two types of processes: SM particle decays (more specifically, Higgs and $Z$ boson decays) and limits coming from direct detection, namely, from the LZ experiment. Turning on one Wilson coefficient at a time, we studied the regions in parameter space which are experimentally excluded. In order to show the artificially large bounds one would obtain without considering the improved power counting, we compare the limits obtained using this rescaling of the field with the so-called ``naive'' power counting, in which the Wilson coefficient has simply the $M_\star$ dependence dictated by the dimensions of the operator. The difference is particularly relevant when we consider high energy observables (Higgs and $Z$ decays): with the naive power counting the bounds grow larger and larger the smaller the DM mass is taken, while this effect is correctly avoided once the improved power counting is considered. At low energy (direct detection) we also observe interesting effects, as strong limits are obtained on two types of operators: the one typically related to spin-independent direct detection experiments that sees the coherent enhancement and those obtained by the exchange in the t-channel of a massless photon, which are enhanced (rather than suppressed) by the small momentum exchange. This second class of operators are the spin-1 analog of electric and magnetic dipole operators that appear for spin-1/2 DM. However we emphasize that complex vectors are the only one that can display very large electric/magnetic dipole moments. 

Finally, we turned to the question: which kind of UV completions can generate the EFT we have constructed? We presented two theories in which $V_\mu$ emerges as a gauge boson.
In both cases, we have shown explicitly how the improved power counting is obtained once heavy dark states are integrated out. Since the relevance of the bounds crucially depends on the power counting, it could be interesting in the future to explore more generic schemes that allows for an even bigger enhancement of the operators constructed with SM field strength and complex vector DM fields, even though the UV completion might be difficult to identify.

\subsubsection*{Acknowledgments}
The work of EB is partly supported by the Italian INFN program on Theoretical Astroparticle
Physics (TAsP), by ``Funda\c{c}\~ao de Amparo \`a Pesquisa do Estado de S\~ao Paulo”
(FAPESP) under contract 2019/04837-9, as well as by Brazilian “Conselho
Nacional de Deselvolvimento Cient\'ifico e Tecnol\'ogico” (CNPq). 
AT acknowledges partial support from MIUR grants PRIN No. 2017L5W2PT.

\appendix

\section{Polarizations of massive vectors}\label{app:polarizations}
\subsection*{Definitions and properties}
In this section we collect useful formulas for the computation of matrix elements in the EFT of complex dark photon. The free Lagrangian of our massive vector can be read out from the first two terms of eq.\,\eqref{eq:basic}. The equations of motion are given by
\be
\partial_\mu V^{\mu\nu} + M^2 V^\nu = \partial^\nu(\partial_\mu V^\mu)\, ,
\ee
and correspond to the Klein Gordon equation when the condition $\partial_\mu V^\mu=0$ is imposed upon the field. This selects three physical polarizations for $V_\mu$, which we will denote by $\varepsilon_\mu^s(p)$. Going to Fourier space and writing the 4-momentum of the particle as $p^\mu=(E_p,\vec p)$, we have that the polarization vectors satisfy $p^\mu \varepsilon_\mu^s=0$ and $g_{\mu\nu}\varepsilon^\mu_s \varepsilon^\nu_{s'}=-\delta_{ss'}$, where as usual the index $s$ can take values $s = \pm$ for the transverse degrees of freedom and $s=L$ for the longitudinal one. They can be constructed from $\vec p$ as follows:
\be
\varepsilon_{\pm}^\mu=(0,\vec{\epsilon}_\pm) , \quad \quad \varepsilon_L^\mu=\bigg(\frac{|\vec p|}{M},  \frac{E_p}{M} \frac{\vec p}{|\vec p|}\bigg)\,,
\ee
where $\vec \epsilon_\pm \cdot \vec p=0$. A useful way of rewriting the above polarization vectors, suitable for taking the non-relativistic limit, can be found as follows. In the rest frame of the particle we have $k^\mu=(M,\vec 0)$, and in that frame the three polarizations (that can be chosen to be eigenvectors of rotations around an axis, conventionally taken to be $z$) are in the fundamental representation of the little group SO(3). We can thus write $\varepsilon^\mu_s(k)=(0,\xi^i_s)$ and the 3-polarizations can be taken to be orthonormal, $\xi^\dag_s \xi_t=\delta_{st}$, with $s,t=1,2,3$. In a generic frame in which the particle has 4-momentum $p^\mu=(E_p,\vec p)$, the polarization vectors are obtained as $\varepsilon_s^\mu(p)=D[L(p,k)]^\mu_{\phantom{\mu}\nu} \varepsilon_s^\nu(k)$, where $L(p,k)$ is the Lorentz transformation on $k$ such that $p^\mu= L(p,k)^\mu_{\,\,\nu}k^\nu$, while $D[L]$ is a suitable representation of $L$. For vectors $D[L]=L$, since they transform in the 4-dimensional representation. Therefore, the polarization in a generic frame can be written as
\be\label{eq:eps-2}
\varepsilon^\mu_s(p)=\bigg(\frac{\vec p \cdot \vec\xi_s}{M}, \vec\xi_s + \frac{(\vec p \cdot \vec\xi_s)}{M(E_p+M)} \vec{p}\bigg)\,,
\ee
which clearly satisfies the relativistic condition $p_\mu \varepsilon^\mu_s(p)=0$ for any value of $\vec p \cdot \vec \xi_s$.
In particular, in the non-relativistic limit we can neglect  terms of order $p^2/M^2$, obtaining
\be
\varepsilon^\mu_s(p)|_{\rm NR}=\bigg(\frac{\vec p \cdot \vec\xi_s}{M}, \vec\xi_s\bigg)+ \mathcal{O}\left(\frac{\vec{p}\,^2}{M^2}\right).
\ee
In presence of two particles we have also $p'=p-q$
\be
\varepsilon^\mu_s(p')|_{\rm NR}=\bigg(\frac{\vec p\,' \cdot \vec\xi_{s'}}{M}, \vec\xi_{s'}\bigg)+\cdots
\ee
The spin operator $J_\sigma(\hat n)$ is well-defined around any generic axis $\hat n$ and the 3-vectors $\vec{\xi}$ and $\vec{\xi}'$ are its eigenstates; more precisely, $J_\sigma(\hat n)\xi_s=\sigma \xi_s$, with $\sigma=-1,0,1$. 

It is important to observe that, since the structure $\bar{\xi}^a_{s'}\xi^b_{s}$, that will appear in the scattering amplitudes, is obtained by the product of two SO(3) fundamentals, it can as usual be decomposed under SO(3) as $\bm{3} \times \bm{3} = \bm{5} + \bm{3}_A + \bm{1}$, with $\bm{5}$ symmetric and traceless, $\bm{3}_A$ antisymmetric and $\bm{1}$ a singlet (trace),
Therefore we have 
\be
\bar{\xi}^a_{s'}\xi^b_{s}=\Sss^{ab} -\frac{i}{2}\epsilon^{abc} S^c_{s's} +\frac13\delta^{ab}\IV \,.
\ee
The appearance of a symmetric tensor structure is typical of a spin-1 DM candidate and is not present for spin-1/2 candidates, since group theory dictates that $\bm{2} \times \bm{2} = \bm{3}_S + \bm{1}_A$. 

In section 6.1 of ref.\,\cite{Gondolo:2020wge} a similar decomposition is carried out, without singling out the trace part from the symmetric combination. We stress, however, that our decomposition follows directly from the irreducible spin representations of the product of polarization vectors of massive vectors and is thus more natural from this point of view.

\subsection*{Massive vectors polarization bilinears}\label{app:pol bilinears}
As mentioned above, the product $[\bar{\xi}^a_{s'}\xi^b_{s}]$ appears in the matrix elements of the elastic DM-nucleus scattering. In particular, the possible structures are the ones generated by the operators $\OS$, $\Op_{FS}$, $\Op_{PS}$, $J_\mu^{V,P}$ and $\mathcal{O}_{\mu\nu}^{A,S,T}$ listed in table \ref{table:structures}. To fix our notation, we will take the incoming DM particle to have momentum $p^\mu$ and spin state $\xi_s$, while the outgoing DM particle has momentum $p'$ and spin state $\xi_{s'}$:
\be
\mathrm{DM}(\xi_{s},\vec p) + \mathrm{SM}(\vec k,\cdots) \to \mathrm{DM}(\xi_{s'},\vec p\,') + \mathrm{SM}'(\vec k',\cdots)\,,
\ee
where DM is our complex dark photon $V$. We also remind the set of Galilean invariant quantities of eq. \eqref{eq:NREFT_structures} which will be used in the following.
We summarize in table \ref{tab:polbilinears} the non-relativistic expansion for the polarization vector bilinears generated by such interactions, where we stop the NR expansion at leading order in a $q^n$ expansion. The few the extra terms appearing in tensor structures components are kept since they would induce NR contributions that are NLO for the single UV operator but of the same order of other effects at the level of NREFT. In fact, for instance, terms like $q^2(\Op_{1,19}^{\textup{\tiny{NR}}})$ that arise from the dipole-like interactions when $\Op_{\mu\nu}^A$ is contracted with the hyper-charge field strength, are clearly NLO in a momentum expansion due to the $\propto q^2$ factor, but would be competing effects of the same order of similar contributions from other effective operators once we take into account the long-range enhancement. 

On top of that, we provide all the useful tools for a more complete analysis, but we urge the reader to remind that we have been coherent with our purpose of investigating only the characteristic LO signature of each UV Wilson coefficient when discussing direct detection phenomenology.

\begin{table}[tb]
\centering
{\def\arraystretch{1.5}
\begin{tabular}{|l ||l |}
\hline
\multicolumn{1}{|c||}{Structure} & \multicolumn{1}{c|}{Polarization Bilinear}\\
\hline\hline
\multicolumn{2}{|c|}{\bf{Scalar}} \\
\hline
$O_S=V_\mu\bar V^\mu$ & $-\IV+\cdots$ \\
$O_{FS}=V_{\mu\nu}\bar V^{\mu\nu}$ & $-2M^2\IV+\cdots$ \\
$O_{PS}=\varepsilon^{\mu\nu\rho\sigma}V_{\mu\nu}\bar V_{\rho\sigma}$ & $-4M(i\vec q\cdot\SV)+\cdots$ \\
\hline\hline
\multicolumn{2}{|c|}{\bf{Vector}} \\
\hline
$J_\mu^V=V_\nu\overset{\leftrightarrow}{\partial_\mu}\bar V^\nu$ & $
\begin{cases} 
 -2M\, \IV +\cdots& (\mu=0)\\
 +P_i\,\IV +\cdots& (\mu=i) 
\end{cases}$\\
$J_\mu^P=\varepsilon_{\mu\nu\rho\sigma}V_\nu\overset{\leftrightarrow}{\partial^\rho}\bar V^\sigma$& $\begin{cases} 
 -i\vec S_V\cdot\vec P +\cdots& (\mu=0)\\
 +2iMS_{V,\,i} +\cdots& (\mu=i) 
\end{cases}$\\
\hline\hline
\multicolumn{2}{|c|}{\bf{Tensor}} \\
\hline
$O^A_{\mu\nu}=V_{[\mu}\bar V_{\nu]}$ & $
\begin{cases} 
 0& (\mu=0,\nu=0)\\
 -\frac{1}{M}\left(q_a\Sss^{ai}-\frac{i}{2} \epsilon_{iac}P^a S_V^c+q_a\frac{\delta^{ai}}{3}\IV\right)+\cdots& (\mu=0,\nu=i)\\
 i\varepsilon_{ijk}S_{V}^k+\cdots&(\mu=i,\nu=j)\\
\end{cases}$ \\
$O^S_{\mu\nu}=V_{(\mu}\bar V_{\nu)}$ & $
\begin{cases}
\frac{1}{2M^2}\SV\cdot(i\vec q\times\vec P)+\dots& (\mu=0,\nu=0) \\
-\frac {1}{M}\left(P_a\Sss^{ai}+\frac i2\epsilon_{iac}q^aS_V^c+\frac{\delta^{ai}}{3}P_a\IV\right)+\dots& (\mu=0,\nu=i)\\
2\left(\Sss^{ij}+\frac{\delta^{ij}}{3}\IV\right)+\dots&(\mu=i,\nu=j)\\
\end{cases}$ \\
 $O^T_{\mu\nu}=V^{}_{[\mu\rho}\bar V^\rho_{\nu]}$ & $
\begin{cases}
 0& (\mu=0,\nu=0) \\
 -iM\varepsilon_{iab}q^aS_V^b+\cdots& (\mu=0,\nu=i)\\
 iM^2 \varepsilon_{ijk}S_V^k +\cdots&(\mu=i,\nu=j) \\
\end{cases}$\\
\hline
\end{tabular}
}
\caption{We summarize the full set of polarization bilinears generated by each operator structure appearing in our EFT for a massive complex vector DM. We classified the structures of bilinears accordingly, making clear the different components for vector and tensor structure.}
\label{tab:polbilinears}
\end{table}

\subsection*{Sum rules}\label{app:sumrules}
In computing the squared amplitudes we often encounter the sum over the spin indices $s$ and $s'$ of initial states. We here list the relevant sum rules for spin-1 particles:
\begin{align}\begin{aligned}
 & \sum_{s's}\vec A\IV  =3A, & ~~~~ &\sum_{s's}A_a\Sss^{ab}S_{s's}^b  =0 , \\
& \sum_{s's}A_a S^a_{s's}  =0 & ~~~~ & \sum_{s's}A_a\Sss^{ab}B_bC_cS_{s's}^c  =0 , \\
&\sum_{s's}(\vec S_{s's})^2 =-6 &~~~~ &\sum_{s's}A_a\Sss^{ab}B_b\delta_{s's} =0, \\
& \sum_{s's}\varepsilon_{abc}\varepsilon_{ajk}A^bS_{s's}^cB^jS_{s's}^k  =-4\vec A\cdot\vec B , & ~~~~ &\sum_{s's}A_a\Sss^{ab}B_c\Sss^{cb}  =\frac 53 \vec A\cdot \vec B , \\
& \sum_{s's}A_aS_{s's}^aB_bS_{s's}^b  =-2\vec A\cdot\vec B, & ~~~~&\sum_{s's}A_a\Sss^{ab}B_bA_c\Sss^{cd}B_d =\frac 12 A^2 B^2+\frac 16 (\vec A\cdot \vec B)^2.
\end{aligned}
\end{align}

\section{Cross sections for direct detection}\label{app:xsectionDD}
We collect here useful equations to compute the rate of events at direct detection experiments. Given the NREFT of eq.\,\eqref{eq:NREFT}, the squared amplitude at the nucleon level, averaged over DM and nucleon spin degrees of freedom, can be written as
\be
\overline{|\M_N|^2}=\frac{1}{2S_V+1}\frac{1}{2S_N+1}\sum_{s',s}\sum_{r',r}\sum_{i,j}c_i^Nc_j^{N'}\bra{N_{s'} V_{r'}}\Op^{\tiny \rm NR}_i\Op^{\tiny \rm NR}_j\ket{N_s V_r},
\ee
where $r,r',s,s'$ are spin indices. From this matrix element we can compute the DM-nucleon scattering cross section as
\be
\sigma_N=\frac{\mu_N^2}{16\pi M^2m_N^2}\overline{|\M_N|^2} .
\label{eq:sigmaN}
\ee
In order to compute the squared matrix element for the DM-target nucleus scattering amplitude, we simply need to multiply by the normalization factor $m_T^2/m_N^2$ (where $m_T$ is the target nucleus mass) that accounts for the $\ket{N} \to \ket{T}$ substitution and to replace the $\sum_{s',s}\sum_{r',r}\bra{N_{s'} V_{r'}}\Op^{\tiny \rm NR}_i\Op^{\tiny \rm NR}_j\ket{N_s V_r}/(2 S_V+1)(2S_N+1)$ matrix element by a nuclear response function $F_{ij}^{NN'}(q^2,v^2)$\,\cite{Fitzpatrick:2012ix,Cirelli:2013ufw,DelNobile:2021wmp} . We obtain
\be
\overline{|\M_T|^2}=\frac{1}{2S_V+1}\frac{1}{2S_T+1}\sum_N\sum_{s',s}\sum_{r',r}|\M_N|^2=\frac{m_T^2}{m_N^2}\sum_N\sum_{i,j} c_i^Nc_j^{N'} F_{ij}^{NN'}(q^2,v^2) .
\label{eq:MT}
\ee
The nuclear response functions account for the coherent enhancement of the cross section at low exchanged momentum and are different for different NR interactions. Moreover, they depend on the kinematic variables $q^2$ and $v^2$, where $v$ is the DM velocity. For convenience, we report here the relevant expressions:
\begin{align}\label{eq:nuclear form factors}
\begin{aligned}
F_{1,1}^{N'N}(M,q^2,v^2) & =F_M^{N'N}=\mathcal{N}_{N'}\mathcal{N}_NF^2_\textup{SI}(E_R)\qquad \text{with }\mathcal{N}_p=Z\,,\,\mathcal{N}_n=A-Z\,,\\
F_{5,5}^{N'N}(M,q^2,v^2) &= \frac{C(S_X)}{4}\frac{1}{q^4}\bigg[q^2\bigg(v^2-\frac{q^2}{4\mu_T^2}\bigg)F_M^{N'N}\bigg]\, ,\\
F_{8,8}^{N'N}(M,q^2,v^2) & =\frac{C(S_X)}{4}\bigg(v^2-\frac{q^2}{4\mu_T^2}\bigg)F_M^{N'N}\, ,\\
F_{11,11}^{N'N}(M,q^2,v^2) &= \frac{C(S_X)}{4}q^2F_M^{N'N}\, ,\\
F_{17,17}^{N'N}(M,q^2,v^2) & = \frac{C(S_X)}{4}q^2\bigg(v^2-\frac{q^2}{4\mu_T^2}\bigg)F_M^{N'N}\, ,\\
F_{19,19}^{N'N}(M,q^2,v^2) & = \frac{C(S_X)}{4}q^4F_M^{N'N}\, ,
\end{aligned}
\end{align}
where the DM spin dependent factor $C(S) = 4 S (S+1)/3 = 8/3$ for a spin-1 candidate. No interference arise between different operators. 
Approximated expressions for $F_{M}^{N'N}$ are listed for different target nuclei in Ref.\,\cite{Fitzpatrick:2012ix}. For simplicity, in our analysis we use the Helm form factor $F_{\rm SI}^2(E_R)$, whose expression can be found in\,\cite{Fitzpatrick:2012ix,Cirelli:2013ufw,DelNobile:2021wmp}. This approximation does not introduce any large error, since at very low recoil energy ({\it i.e.} in the most relevant region for direct detection experiments) $F_M$ and $F_{\rm SI}^2$ give basically the same results. 

The differential DM-target scattering cross section in the LAB frame is given by:
\be
\frac{d\sigma_T}{dE_R}=\frac{\overline{|\M_T|^2}}{32\pi M^2m_Tv^2} .
\label{eq:dsdER}
\ee
Finally, we are in the position of giving explicit expressions for the differential detection rate\,\cite{Cirelli:2013ufw,DelNobile:2021wmp}:
\be
\frac{dR}{dE_R}(E_R) =\frac{\rho_\textup{DM}}{32\pi M^3m_N^2}A^2\sum_{i,j}c_ic_j\mathscr{F}_{i,j}^{N'N}(M,E_R),
\label{eq:dRdER}
\ee
where $\rho_\textup{DM}$ is the DM density on Earth and the integrated form factors expressions are given by
\be
\mathscr{F}_{i,j}^{N'N}(M,E_R)=\int_{v\geq v_\textup{min}(E_R)}d^3v \frac{f_\textup{SHM,E}}{v}(\vec v)F_{i,j}^{N'N}(M,q^2,v^2)\bigg|_{q^2=2m_TE_R}.
\label{eq:integrated F}
\ee
In the previous equation, we take the velocity distribution $f_{\rm SHM,E}$ in Earth's frame to be
\be
f_{\rm SHM,E} = \frac{e^{-(v+v_E)^2/v_0^2}-\beta e^{-v_{\tiny{esc}}^2/v_0^2}}{(v_0 \sqrt{\pi})^3N_{\tiny{esc}}}\Theta(v_{\tiny{esc}}-|v+v_E|)\,,
\ee
with $N_{\tiny{esc}}$ being a normalization factor which goes to one as we send the escape velocity to infinity. The minimum velocity for which we have a direct detection scattering event is $v_{min}(E_R) = (m_T E_R/2 \mu_{TV}^2)^{1/2}$, with $\mu_{TV} = M m_T/(M+m_T)$ the DM-target reduced mass. From the practical point of view, we compute the velocity integrals using the expressions presented in Appendix A of \cite{Cirelli:2013ufw}. The expected numbers of observed events, computed as a function of the EFT parameters $(\wc_i, M_\star, g_\star)$ can be found integrating over the nuclear recoil energy in the range indicated by experiments, after having multiplied eq.\,\eqref{eq:dRdER} by the total exposure:
\be
N_\textup{th} =T_\textup{exp}M_\textup{exp}\int_{E_{R_\textup{min}}}^{E_{R_\textup{max}}}\frac{dR}{dE_R}(E_R),
\label{eq:Nth}
\ee
where $T_\textrm{exp}$ is the total time over which the experiment ran and $M_\textrm{exp}$ the total mass of the experiment.

\pagestyle{plain}
\bibliographystyle{jhep}
\small
\bibliography{biblio-complex}

\end{document}